\documentclass[11pt]{article}
\usepackage[letterpaper,margin=1in]{geometry}
\usepackage[T1]{fontenc}
\usepackage{lmodern}
\usepackage{amsmath,amssymb,amsthm,mathtools}
\usepackage{array}
\usepackage{booktabs}
\usepackage{tabularx}
\usepackage{microtype}
\usepackage{tikz-cd}
\usepackage{tikz}
\usetikzlibrary{arrows.meta,calc}
\usepackage{tikz}
\usetikzlibrary{arrows.meta,positioning}
\usepackage{algorithm}
\usepackage{float}
\usepackage{algpseudocode}
\usepackage{fancyhdr}
\usepackage[hidelinks]{hyperref}
\usepackage{comment}

\numberwithin{equation}{section}

\newtheorem{theorem}{Theorem}[section]
\newtheorem{lemma}[theorem]{Lemma}
\newtheorem{proposition}[theorem]{Proposition}
\newtheorem{corollary}[theorem]{Corollary}
\theoremstyle{definition}
\newtheorem{definition}[theorem]{Definition}
\theoremstyle{remark}

\theoremstyle{plain}

\DeclareMathOperator{\GL}{GL}
\DeclareMathOperator{\SL}{SL}
\DeclareMathOperator{\Mat}{Mat}
\DeclareMathOperator{\Disc}{Disc}

\DeclareMathOperator{\Jac}{Jac}

\DeclareMathOperator{\covol}{covol}
\DeclareMathOperator{\im}{im}
\DeclareMathOperator{\rank}{rank}
\DeclareMathOperator{\spanQ}{span_{\mathbb Q}}
\DeclareMathOperator{\diag}{diag}
\DeclareMathOperator{\coker}{coker}

\newcommand{\Z}{\mathbb Z}
\newcommand{\Q}{\mathbb Q}

\newcommand{\tensor}{\otimes}

\newcommand{\strictiso}{\sim_{\mathrm{str}}}

\newcommand{\transpose}{\mathsf T}

\algrenewcommand\algorithmicrequire{\textbf{Require:}}
\algrenewcommand\algorithmicensure{\textbf{Ensure:}}
\algrenewcommand\algorithmicthen{\textbf{then}}
\algrenewcommand\algorithmicdo{\textbf{do}}
\algrenewcommand\algorithmicend{\textbf{end}}
\algrenewcommand\algorithmicif{\textbf{if}}
\algrenewcommand\algorithmicelse{\textbf{else}}
\algrenewcommand\algorithmicfor{\textbf{for}}
\algrenewcommand\algorithmicreturn{\textbf{return}}

\usepackage{xspace}
\newcommand{\BQP}{\ensuremath{\mathsf{BQP}}\xspace}
\newcommand{\AM}{\ensuremath{\mathsf{AM}}\xspace}
\newcommand{\CH}{\ensuremath{\mathsf{CH}}\xspace}

\hypersetup{
  pdftitle={Integral Tensor Isomorphism via Cubic Rings and Lattice Methods},
  pdfsubject={Integral tensor isomorphism, balanced modules, and quantum algorithms},
  pdfkeywords={tensor isomorphism, Bhargava, Wood, balanced modules, lattices, BQP},
  pdfauthor={}
}

\title{Integral Tensor Isomorphism via Cubic Rings and Lattice Methods}
\author{
    Thomas Huffstutler \\
    \small Department of Computer Science, \\
    \small Stony Brook University \\
    \small \texttt{thuffstutler@cs.stonybrook.edu}
}
\date{}
\begin{document}
\maketitle

\begin{abstract}
    We prove that isomorphism of integral \(2\times3\times3\) tensors under \(\mathrm{GL}_2(\mathbb{Z})\times\mathrm{GL}_3(\mathbb{Z})\times\mathrm{GL}_3(\mathbb{Z})\) is decidable in bounded-error quantum polynomial time. For a tensor \(T=(A,B)\) with nonzero determinant form $\det(xA-yB)$, we use the parametrization of Wood \cite{wood} by balanced modules to reduce the isomorphism problem to principality tests for compatible lattices, and use the algorithms of Bley, Hofmann, and Johnston \cite{bhj} to solve this in $\BQP$. When the determinant form vanishes identically, we get either an integral $2\times2\times2$ core, or a polynomial kernel vector of degree one or two. The linear-kernel case yields another compatible lattice problem, while the quadratic-kernel case reduces to a fixed system of polynomial congruences and unimodular lifting. A quantifier-elimination argument decides these congruences in deterministic polynomial time given the prime factorization of the modulus, which quantum integer factorization supplies. Finally, we show that zero padding preserves integral $2\times2\times2$ tensor isomorphism, yielding a polynomial-time reduction from real-quadratic principality to the identically singular $2\times3\times3$ problem.
\end{abstract}

\tableofcontents

\newpage

\clearpage

\section{Introduction}\label{sec:introduction}

Bhargava's seminal works on \emph{Higher Composition Laws}, revealed a surprising arithmetic structure in tensors with integer entries, giving a direct correspondence between their orbits and natural number theoretic objects. In the case, of $2\times3\times 3$ tensors, a nondegenerate pair of $3\times3$ integer matrices encodes a cubic ring together with a balanced pair of fractional ideals~\cite{bhargava}, connecting the characterization of these tensors under the action of $\SL_{2}(\mathbb{Z})\times\SL_{3}(\mathbb{Z})\times\SL_{3}(\mathbb{Z})$ to questions about ideals, units, and integral lattices. This unexpected association also provides an opportunity to understand the structure of such tensors algorithmically. Several of the arithmetic problems that arise from such correspondences admit quantum polynomial-time algorithms, even when polynomial-time classical algorithms are not known~\cite{kitaev,BiasseSong, hallgren_pit}. Thus, a natural question is whether we can assemble these arithmetic algorithms into efficient algorithms to test the original tensor orbits.

Grochow, Qiao, Stange, and Sun used Bhargava's \emph{Higher Composition Laws I}~\cite{bhargava2004higher1} to show that the isomorphism problem for orbits of $2\times2\times 2$ tensors isomorphism is polynomial-time equivalent to a quadratic principal ideal testing problem and therefore belongs to \BQP~\cite{gqss}, thus turning a composition law into an algorithmic bridge. The next natural composition law, Bhargava's \emph{Higher Composition Laws II} concerning $2\times3\times3$ tensors, presents additional algorithmic difficulties for the isomorphism testing problem due to the number theoretic objects associate with such tensors, namely, a pair of modules and cubic ring~\cite{wood}. The relevant integral modules for two tensors must be compared according to a certain balancing condition between them, and in addition, the determinant form can have repeated roots or vanish identically. Both of these additional structures contribute to the reason that there is not a straightforward modification to the principal ideal testing reduction and corresponding algorithm from the $2\times 2\times 2$ case, that allows us to adapt it to the $2\times 3\times 3$ case.

\subsection{Our results}\label{subsec:our-results}

Our main result shows that we can indeed generalize the results of \cite{gqss} to show that quantum algorithms are able to efficiently decide whether two $2\times 3\times 3$ tensors are isomorphic.

\begin{theorem}[Integral tensor isomorphism]\label{thm:main}
There is a bounded-error quantum polynomial-time algorithm that, given
$T,T'\in\Z^2\tensor\Z^3\tensor\Z^3$, decides whether they are isomorphic under
$\GL_2(\Z)\times\GL_3(\Z)\times\GL_3(\Z)$.
\end{theorem}

The main invariant organizing the proof is the binary cubic
\[
        f_T(x,y)=\det(xA-yB).
\]
which we use to split the proof up into essentially two cases, when $f_{T}\not\equiv 0$ and $f_{T}\equiv0$.

Our next result concerns compatible balanced-module pair comparison using Wood's extension of Bhargava's parametrization, which associates to a nonzero cubic $f$, a cubic ring $R_{f}$ and a pair of modules that are free of rank three as abelian groups, and equipped with a balancing map~\cite{wood}. Any isomorphism must therefore preserve the pairing and its balancing map, in addition to the module actions. When considering a fixed binary cubic, we obtain the following consequence en-route to our main theorem.

\begin{theorem}[Balanced pairs for a nonzero cubic]\label{thm:intro-balanced}
Given a nonzero integral binary cubic $f$ and two balanced pairs of modules for $f$, each module free of rank three over $\Z$ and specified in an integral basis with its ring action and balancing pairing, isomorphism of the balanced pairs over the ring $R_f$ can be decided in \BQP. Equivalently, strict equivalence of two integral $2\times3\times3$ tensors with the same nonzero determinant form is in \BQP.
\end{theorem}

Here \emph{strict equivalence}, denoted $T\strictiso T'$, fixes the first $\GL_{2}(\Z)$ factor and uses row and column transformations with equal determinants. Proposition~\ref{prop:fixed-form} proves the fixed-form comparison, and Theorem~\ref{thm:wood} identifies it with the balanced-pair problem. Section~\ref{subsec:overview-principality} explains the decision criterion.

We additionally identify several families of problems within the larger main problem that admit classical algorithms. A nonzero cubic is a \emph{pure cube} if it has the form $C\ell^3$, where $C$ is a nonzero integer and $\ell$ is a primitive integral linear form. Its stabilizer contains infinitely many integral shears, so a finite enumeration of first-factor transformations is unavailable. Nevertheless, canonical subspaces of an associated nilpotent operator give a classical algorithm.

\begin{theorem}[Classical subfamilies]\label{thm:intro-classical}
Integral $2\times3\times3$ tensor isomorphism is decidable in deterministic polynomial time on each of the following invariant families: tensors with dependent slices; regular tensors whose determinant form is a pure cube; and tensors of normal rank at most one.
\end{theorem}

The normal rank is the rank of $xA-yB$ over $\Q(x,y)$. The first and third families reduce to Smith normal form (Lemmas~\ref{lem:dependent-slices} and~\ref{lem:low-normal-rank}), which is computable in deterministic polynomial time~\cite{kannan-bachem}. The pure-cube family requires a separate argument where a canonical line inside a canonical plane reduces the unbounded shear parameter to an integer linear system and one congruence. Corollary~\ref{cor:pure-cube} proves the pure-cube bound. Sections~\ref{subsec:overview-first-factor} and~\ref{subsec:overview-singular} explain these reductions.

When $f_T\equiv0$, no slice combination is invertible and the nonzero-form parametrization of ~\cite{wood} does not apply. Instead, we analyze the polynomial kernel of the matrix pencil $M_{T}(x,y)=xA-yB$ through invariant subspaces, and compatible lattice data. Their comparison requires the following auxiliary result which may be of interest independent of its particular application here, thus we state it explicitly.

\begin{theorem}[Fixed-format polynomial congruences]\label{thm:intro-congruences}
Fix positive integers $n,m,d$. Given $m$ integer polynomials of degree at most $d$ in $n$ variables, and an integer $N\ge2$ together with its prime factorization, one can decide in deterministic polynomial time in the binary input length whether the polynomials have a common zero modulo $N$.
\end{theorem}

Lemma~\ref{lem:fixed-format-congruences} proves this consequence of established quantifier-elimination results~\cite{pas,macintyre,Kie76,Coo72}.

\subsection{Relation to previous work}\label{subsec:related-work}

\paragraph{Tensor isomorphism over fields and quantum information.}
Tensor isomorphism has a direct interpretation in the classification of entanglement. A pure state shared by several parties is a complex valued tensor whose factor dimensions are the dimensions of their local Hilbert spaces. Two pure states are equivalent under stochastic local operations and classical communication (SLOCC) precisely when invertible complex maps on the individual factors relate them, with normalization absorbed into the maps~\cite{DVC00}. Thus a $2\times3\times3$ tensor describes one qubit and two qutrits. Chitambar, Miller, and Shi use matrix pencils to decide SLOCC equivalence for $2\otimes m\otimes n$ systems~\cite{CMS10}. Restricting both states and allowed local operations to the reals leads to a different quantum theory and different entanglement questions~\cite{CFR01}. Integral equivalence retains still finer data: a nonzero integral tensor $T$ and $2T$ define the same normalized complex state, but cannot be integrally isomorphic because unimodular transformations preserve the greatest common divisor of the entries.

\paragraph{Integral orbits and the role of the base ring.}
Integral tensor isomorphism is an orbit problem for an arithmetic group. The algorithms of Grunewald and Segal establish decidability for such orbit questions in the relevant rational representations~\cite{grunewald-segal}, but decidability does not provide the polynomial bit-complexity bound sought here. The recent complexity-theoretic study of tensor isomorphism over commutative rings by Grochow, Qiao, Stange, and Sun~\cite{gqss} makes these quantitative questions explicit and provides the closest algorithmic predecessor to our theorem.

For binary-encoded rational-entry tensors, equivalence under complex changes of basis can be expressed by polynomially many equations~\cite{GGPS23}. Combining this encoding with the result of Andrews, Garg, and Schost on Hilbert's Nullstellensatz places this version of tensor isomorphism unconditionally in the counting hierarchy, \CH~\cite{AGS26}, and the seminal result of Koiran gives membership in \AM under the generalized Riemann hypothesis~\cite{Koi96}. These bounds concern complex solutions, whereas an integral isomorphism must satisfy the additional lattice and unimodularity conditions that our algorithm addresses.

\paragraph{Composition laws and balanced pairs.}
Bhargava's \emph{Higher Composition Laws II} describes nondegenerate $2\times3\times3$ tensors using cubic rings and balanced ideal pairs~\cite{bhargava}. Wood extends the description to balanced pairs of modules for every nonzero binary form~\cite{wood}, removing the nondegeneracy condition. We use Wood's characterization of nonzero binary cubics to obtain a single algorithm across all discriminants. We treat the case when $f_{T}\equiv 0$, the identically zero form, separately since this case is not handled in Wood's generalization. We instead handle this case through a direct argument about certain invariant quantities of the polynomial kernel of the matrix pencil $M_{T}$.

\paragraph{Module isomorphism.}
Ordinary module isomorphism has substantial existing algorithmic theory. Chistov, Ivanyos, and Karpinski give classical polynomial-time algorithms for explicitly represented modules over finite-dimensional algebras over finite fields or algebraic number fields~\cite{chistov-ivanyos-karpinski}. Over integral orders, Bley, Hofmann, and Johnston reduce lattice isomorphism to principality in an endomorphism order, and give a quantum polynomial-time reduction to the required explicit semisimple decomposition under their splitting hypothesis~\cite{bhj}. They also prove a quantum polynomial-time algorithm for integral matrix similarity~\cite{bhj}.

For an individual module in our cubic setting, clearing denominators in the action of a generator of $R_f\tensor_{\Z}\Q$ already reduces module comparison to that integral matrix-similarity problem~\cite{bhj}. The extra requirement for a balanced pair is that the two module isomorphisms preserve the pairing. We encode this requirement by a single compatible homomorphism lattice and verify the algebraic hypotheses needed by the existing arithmetic algorithms. The same principle reappears on the singular boundary with a different diagram.

\paragraph{Organization.}
Section~\ref{sec:technical-overview} presents the techniques, the complete case division, and the allocation of classical and quantum work. The detailed proofs begin with notation and elementary lattice reductions, followed by the balanced-module comparison and the first-factor action. Sections~\ref{sec:zero-boundary}-\ref{sec:symmetric-comparison} treat singular pencils. Algorithm~\ref{alg:integral-233-ti} assembles the procedure, and Proposition~\ref{prop:faithful-padding} proves the arithmetic reduction. Appendix~\ref{app:bhargava:nondegenerate} gives an alternative proof for nondegenerate tensors using the objects that come directly from Bhargava's \emph{Higher Composition Laws II} \cite{bhargava2004higher1}.

\subsection{Technical overview}\label{sec:technical-overview}
\label{subsec:overview-determinant}

\begin{figure}[H]
\centering
\begin{tikzpicture}[
  box/.style={rounded corners=3pt,line width=0.7pt,align=center,
              font=\small,text width=5.9cm,inner sep=6pt},
  inputbox/.style={box,text width=8cm,minimum height=1.05cm,
                   fill=blue!8,draw=blue!55!black},
  structurebox/.style={box,minimum height=1.65cm,
                       fill=teal!10,draw=teal!55!black},
  comparisonbox/.style={box,minimum height=1.65cm,
                        fill=violet!7,draw=violet!45!black},
  methodbox/.style={box,minimum height=2.65cm,
                    fill=orange!13,draw=orange!60!black},
  arrow/.style={-{Latex[length=2mm]},draw=black!75,line width=0.7pt},
  branch/.style={font=\small,fill=white,inner sep=3pt}
]
\node[inputbox] (input) at (0,0) {
  \textbf{Integral tensor isomorphism}\\
  $T,T'\in\Z^2\tensor\Z^3\tensor\Z^3$
};
\node[structurebox] (modules) at (-3.6,-2.35) {
  \textbf{Balanced modules}\\
  Match determinant forms; retain\\
  two modules and their pairing\\[2pt]
  {\scriptsize Lemma~\ref{lem:first-factor}; Theorem~\ref{thm:wood}}
};
\node[structurebox] (kernels) at (3.6,-2.35) {
  \textbf{Polynomial kernel}\\
  Compute its least degree\\
  $d_T\in\{1,2\}$\\[2pt]
  {\scriptsize Lemma~\ref{lem:polynomial-kernel}}
};
\node[comparisonbox,below=8mm of modules] (pairing) {
  \textbf{Preserve the pairing}\\
  $PA=A'Q,\qquad QD=D'Q$\\
  A compatible integral square\\[2pt]
  {\scriptsize Lemma~\ref{lem:compatible-module-isos}}
};
\node[comparisonbox,below=8mm of kernels] (kerneldata) {
  \textbf{Encode the kernel data}\\
  $d_T=1$: marked diagram $X_T$\\
  $d_T=2$: coefficient matrices $K_T,H_T$\\[2pt]
  {\scriptsize Lemmas~\ref{lem:linear-kernel-diagram}
   and~\ref{lem:quadratic-kernel-data}}
};
\node[methodbox,below=8mm of pairing] (regularmethod) {
  \textbf{Principality and row lattices}\\
  BHJ principality decision\\
  and a Smith-normal-form test\\[2pt]
  {\scriptsize Proposition~\ref{prop:fixed-form}}
};
\node[methodbox,below=8mm of kerneldata] (singularmethod) {
  \textbf{Two arithmetic tests}\\
  $d_T=1$: principality and row lattices\\
  $d_T=2$: fourteen congruences\\
  in four variables; unimodular lifting\\[2pt]
  {\scriptsize Proposition~\ref{prop:linear-kernel-bqp};
   Theorem~\ref{thm:symmetric-bqp}}
};
\draw[arrow] (input.south) --
  node[branch] {$f_T\not\equiv0$} (modules.north);
\draw[arrow] (input.south) --
  node[branch] {$f_T\equiv0$} (kernels.north);
\draw[arrow] (modules.south) -- (pairing.north);
\draw[arrow] (pairing.south) -- (regularmethod.north);
\draw[arrow] (kernels.south) -- (kerneldata.north);
\draw[arrow] (kerneldata.south) -- (singularmethod.north);
\end{tikzpicture}
\caption{The two main proof routes, omitting the elementary cases
and the separate $2\times2\times2$ core comparison.
The nonzero-form route shown is for non-pure cubes; pure cubes use
the separate classical flag test. BHJ denotes
Bley--Hofmann--Johnston~\cite{bhj}. In the quadratic-kernel case,
quantum factorization supplies the input needed by the classical
fixed-format congruence test.}
\label{fig:proof-strategy}
\end{figure}
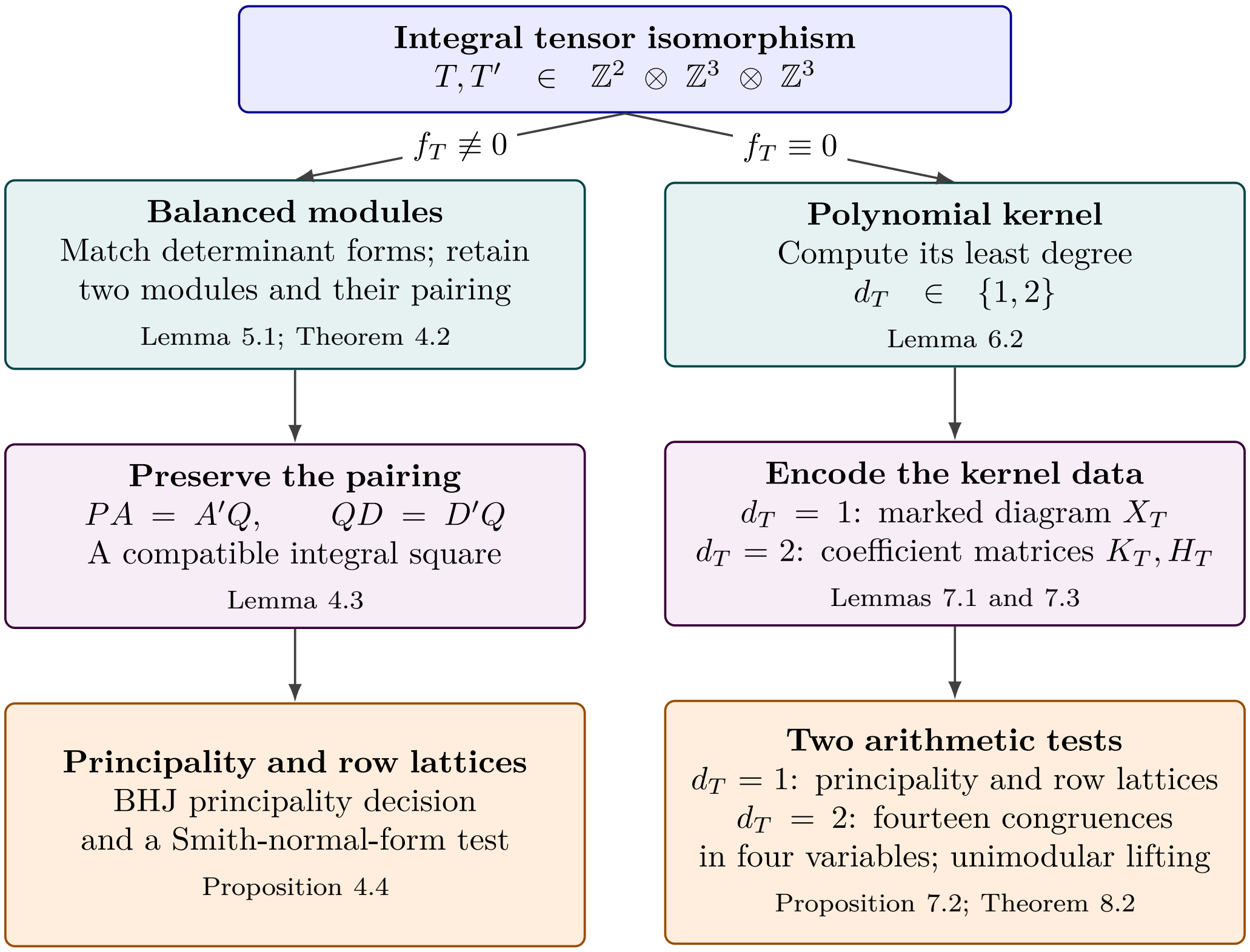

The determinant form divides the proof into two routes. Write
$M_T(x,y)=xA-yB$ for the matrix pencil associated with $T=(A,B)$, and
$f_T=\det M_T$. If $f_T\not\equiv0$, some rational combination of the
slices is invertible and we call this the \emph{regular} case. If
$f_T\equiv0$, every combination is singular. In the regular case, we
retain the integral structure through two modules and a balanced pairing. In the
singular case, we recover it from a vector in the kernel that varies
polynomially with $(x,y)$. Each route has the same progression:
\emph{identify the structure, express the compatibility required of an
isomorphism, and decide that compatibility by an arithmetic test}.
Figure~\ref{fig:proof-strategy} summarizes the routes at the end of this
overview.

We use the isomorphism convention

\[
 PM_{T}(z)=M_{T'}(Sz)Q,\qquad z=(x,y)^{\transpose},
 \qquad S\in\GL_{2}(\Z),\quad P,Q\in\GL_{3}(\Z).
\]

All changes of basis are unimodular, that is, their determinants are $\pm1$.
As a preliminary step, dependent slices reduce to a pair $(A_{0},0)$,
whose comparison is a Smith-normal-form computation
(Lemma~\ref{lem:dependent-slices}). We may therefore assume below that
the slices are independent. Whenever two inputs have different values
of a structural invariant used in the case division, we reject.

\subsubsection{Nonzero determinant: preserve the balanced pair}
\label{subsec:overview-regular}

\paragraph{Match the determinant forms.}\label{subsec:overview-first-factor}
The first factor changes the variables of the binary cubic, while the
other two factors contribute a sign. We consider the regular case where $f_{T}\not\equiv 0$ and spit it into the cases where the discriminant of $f_{T}$ is non-zero vs. zero. In the case where the discriminant is equal to zero, $f_{T}$ must either have a repeated root or a single root with multiplicity three, which we call the \emph{pure cube} case. For a nonzero cubic that is not a pure cube, there are at most six integral changes of variables carrying it to another fixed cubic. Lemma~\ref{lem:first-factor} computes this complete list in polynomial time, using binary cubic equivalence and a separate normalization for a double root~\cite{cremona}. Each candidate leaves a comparison at an exactly fixed form $f_T=f_{T'}=f$. Pure cubes have an infinite stabilizer and we handle that exception at the end of this branch.

\paragraph{Record the pairing by a commuting square.}
\label{subsec:overview-square}
For a fixed nonzero $f$, Wood's parametrization supplies a cubic ring
$R_f$ (a ring free of rank three over $\Z$), two modules $M,N$ that
are free of rank three over $\Z$, and a
pairing $\beta:M\tensor_{R_{f}}N\to R_{f}$ which respects the ring action $\beta(rm,n)=\beta(m,rn)=r\beta(m,n)$~\cite{wood}. Consequently, independently identifying the two
modules does not yet identify the tensor as the chosen maps must preserve
the pairing as well.

Note that some $k\in\{0,1,2,3\}$ has $f(1,k)\ne0$, and replacing both pairs by
$(A-kB,B)$ makes $A,A'$ invertible over $\Q$ with equal determinants.
After this relabeling, let
\[
 D=A^{-1}B,\qquad D'=(A')^{-1}B'.
\]
where the operator $D$ records the second slice relative to the first and
describes the action of a rational generator of the cubic algebra~\cite{wood}.
Fixing an $n\in N$ gives the linear functional $m\mapsto mAn$, and
collecting these functionals turns the pairing into the integral
module map $A:N\to L$, where
$L=\operatorname{Hom}_{\Z}(M,\Z)$ is the dual lattice.
Lemma~\ref{lem:compatible-module-isos} identifies strict equivalence
with the existence of unimodular maps $Q:N\to N'$ and $P:L\to L'$
satisfying
\[
 \underbrace{PA=A'Q}_{\text{preserve the pairing map $\beta$}},
 \qquad
 \underbrace{QD=D'Q}_{\text{respect the module action over $R_{f}$}}.
\]
The first equation is a commuting square. Together the equations recover
the second slice, since $PB=PAD=A'QD=A'D'Q=B'Q$.
When $A=A'=I_3$, the square forces $P=Q$, giving ordinary integral
matrix similarity.

\paragraph{Decide compatibility by principality and a row-lattice test.}
\label{subsec:overview-principality}
We first allow compatible maps to be noninvertible. This makes their
equations linear and gives the homomorphism lattice
\[
 \mathcal H=\{Q\in\Mat_{3}(\Z):QD=D'Q,\ 
                         A'QA^{-1}\in\Mat_{3}(\Z)\}.
\]
Its second condition says that the companion map $P=A'QA^{-1}$ is
integral. Similarly, compatible endomorphisms of the target form
\[
 \mathcal E=\{X\in\Mat_3(\Z):XD'=D'X,\ 
                         A'X(A')^{-1}\in\Mat_{3}(\Z)\}.
\]
Both are computed by integer linear algebra. Composition makes
$\mathcal{E}$ a ring acting on $\mathcal{H}$. The reason to test
\emph{principality} is that if an isomorphism $Q_{0}$ exists,
every compatible map is a target endomorphism composed with $Q_{0}$, and thus $\mathcal H=\mathcal E Q_{0}$.

To put this observation into the arithmetic algorithm, first test
rational similarity of $D,D'$ and, if it succeeds, choose
$J\in\GL_3(\Q)$ with $JD=D'J$. Then
$\mathcal I=\mathcal HJ^{-1}$ lies in the rational centralizer of $D'$,
the algebra of rational matrices commuting with $D'$.
In that algebra, $\mathcal E$ is an \emph{order}, meaning a unital
subring whose additive group is a full integral lattice, and
$\mathcal I$ is a full left $\mathcal E$-lattice.
Principality asks whether $\mathcal I$ is free of rank one over
$\mathcal E$.

Principality alone does not certify an isomorphism since a generator may
have determinant other than $\pm1$. We detect this using the lattice
$\mathcal R(\mathcal H)$ generated by all rows of a $\Z$-basis of
$\mathcal H$. The exact criterion is
\[
 T\strictiso T'
 \quad\Longleftrightarrow\quad
 \mathcal I\text{ is principal over }\mathcal E
 \quad\text{and}\quad
 \mathcal R(\mathcal H)=\Mat_{1 \times 3}(\Z).
\]
Indeed, principality gives a generator with
$\mathcal H=\mathcal E Q_0$. Since every element of $\mathcal E$ is
integral and $I_3\in\mathcal E$, the rows of all compatible maps generate
exactly the row lattice of $Q_0$. Equality with $\Z^{1\times3}$ forces
$Q_0$ to be unimodular and its companion $A'Q_0A^{-1}$ is integral by
construction and has the same determinant, hence is unimodular too.
This proves the criterion without computing the generator.

We therefore need only a principality decision and a Smith-normal-form
test. Bley-Hofmann-Johnston supply the quantum principality
algorithm~\cite{bhj}. The hypothesis of the relevant algorithm is satisfied since the rational canonical form explicitly decomposes the centralizer modulo its nilpotent radical into full matrix algebras over number fields.
This is the decomposition required by the necessary algorithm from ~\cite{bhj}(Proposition~\ref{prop:fixed-form} and~\eqref{eq:semisimple-quotient}).

\paragraph{The pure-cube exception: replace an infinite search by a flag.}
In the case where $f_{T}$ is a pure cube, we normalize it to the form $Cx^{3}$ with $C>0$. Its remaining slice
transformations are
\[
 (A,B)\longmapsto(A+tB,\delta B),
 \qquad t\in\Z,\quad\delta\in\{\pm1\}.
\]
The fixed-form test we applied in the repeated root case still applies, but there are now infinitely many
values of $t$. The useful invariant in this case is a line inside a plane.
Since $D=A^{-1}B$ from ~\cite{wood} is nonzero and $D^3=0$, set
\[
 F_1=\im D\cap\ker D,\qquad F_2=\im D+\ker D.
\]
This nested pair is referred to as a \emph{flag}. The shear replaces $D$ by
$\delta(D-tD^2)$, preserving its image and kernel and hence the flag.
Every isomorphism must therefore carry this flag, and the output flag
$AF_1\subset AF_2$, to their counterparts for the target.

Intersecting the subspaces with $\Z^3$ retains all their integral
points and gives saturated lattices. Bases adapted to these lattices
make $A$ upper triangular and $B$ strictly upper triangular, and force
the comparison maps to be upper triangular. Their diagonal entries
leave only finitely many sign choices. For each choice, the remaining
equations reduce to an integer linear system.
Proposition~\ref{prop:pure-cube-test} makes this calculation explicit
and yields a deterministic polynomial-time algorithm.

\subsubsection{Zero determinant: recover the kernel structure}
\label{subsec:overview-singular}

When $f_T\equiv0$, there is no invertible rational slice and the
nonzero-form parametrization is unavailable, and the replacement is the
kernel of the \emph{pencil} $M_T(x,y)$. We first separate singularity
caused by unused directions, and the way the remaining kernel varies
with $(x,y)$ determines the comparison problem.

\paragraph{Remove the elementary cases.}
The \emph{normal rank} is the rank of $M_T$ over $\Q(x,y)$.
If it is at most one, the independent slices share either an image
line or a row-space line. Removing unused directions leaves a single
integer matrix, handled by Smith normal form
(Lemma~\ref{lem:low-normal-rank}). At normal rank two, compute
\[
 K_R=\ker A\cap\ker B,\qquad
 K_L=\ker A^{\transpose}\cap\ker B^{\transpose}.
\]
If both are nonzero, saturated integral bases expose a common zero
column and a common zero row. Removing them leaves an integral
$2\times2\times2$ tensor, handled by the known quantum
algorithm~\cite{gqss}. In the remaining cases, compare the ordered
pair of common-kernel dimensions and, if necessary, transpose
\emph{both} pencils to arrange $K_R=0$.

\paragraph{Measure how the kernel varies.}
A homogeneous polynomial kernel vector is a nonzero vector
\[
 v_T(x,y)=\sum_{j=0}^{d}x^{d-j}y^jv_j,\quad v_j\in\Q^3,
 \qquad M_T(x,y)v_T(x,y)=0.
\]
Its least possible degree $d_T$ measures how much the kernel direction
must vary with the slice coordinates. Degree zero would be a common
right-kernel vector, which we have excluded. On the other hand, the
cross product of two independent rows of $M_T$ gives a quadratic
kernel vector. Canceling any common polynomial factor therefore leaves
$d_T=1$ or $2$.

At the least degree, the vector is unique up to rational scaling.
Clearing denominators and dividing all integer coefficients by their
gcd gives a primitive integral vector, unique up to sign.
Its coefficients are found by solving
$Av_0=0$, $Av_j=Bv_{j-1}$, and $Bv_d=0$, first for $d=1$ and then,
if needed, for $d=2$ (Lemma~\ref{lem:polynomial-kernel}).
This supplies both an invariant degree and canonical coefficient data.

\paragraph{Degree one: preserve a plane and a line.}
\label{subsec:overview-linear}
For $v_T=xv_0+yv_1$, the relations
$Av_0=0$, $Av_1=Bv_0$, and $Bv_1=0$ determine an input plane
$\spanQ\{v_0,v_1\}$ that both slices send into the output line
$\Q Av_1$. In integral bases adapted to these saturated subspaces,
\[
 M_T(z)=
 \begin{pmatrix}\ell_Tz&(R_Tz)^{\transpose}\\
                 C_Tz&0_{2\times2}\end{pmatrix},
 \qquad
 X_T=\begin{pmatrix}\ell_T\\ C_T\\ R_T\end{pmatrix}:
       \Z^2\longrightarrow\Z^5.
\]
Here $\ell_T$ is a $1\times2$ row, $C_T,R_T$ are $2\times2$
integer matrices, $R_T$ is nonsingular, and $C_T\ne0$.
The five rows of $X_T$ record the five potentially nonzero entries
of the pencil as linear forms.

We must also remember their geometric roles. Mark the sublattices
$\Lambda_1=\Z\oplus\Z^2\oplus0$ and
$\Lambda_2=\Z\oplus0\oplus\Z^2$ in the target $\Z^5$.
Their intersection marks the first coordinate, and their quotients
by that intersection distinguish the $C_T$ and $R_T$ blocks.
Preserving them enforces the block pattern coming from the input
plane and output line. Lemma~\ref{lem:linear-kernel-diagram} gives the
exact criterion
\[
 FX_T=X_{T'}S,\qquad F\Lambda_i=\Lambda_i\ (i=1,2),
 \qquad S\in\GL_2(\Z),\quad F\in\GL_5(\Z).
\]

As in the regular case, we form its homomorphism lattice and target
endomorphism order. After a rational identification, the algorithm of ~\cite{bhj} applies, and
the rational endomorphism algebra has an explicit semisimple quotient
consisting of full matrix algebras over $\Q$.
Principality, together with row-lattice equalities for the $S$ and
$F$ components, certifies an integral isomorphism, and the two equalities
certify unimodularity exactly as before (Proposition~\ref{prop:linear-kernel-bqp}).

\paragraph{Degree two: the kernel coefficients force the other maps.}
\label{subsec:overview-quadratic}
For a quadratic kernel, the three coefficient vectors form a rational
basis of the input space. Write
$v_T(z)=K_T\nu(z)$, where
$\nu(x,y)=(x^2,xy,y^2)^{\transpose}$ and $K_T$ is nonsingular.
The basic linear relations among these monomials are the two rows of
\[
 L(x,y)=\begin{pmatrix}-y&x&0\\0&-y&x\end{pmatrix},
 \qquad L(x,y)\nu(x,y)=0.
\]
The total output span has dimension two. Using an integral basis of
its saturation gives a reduced pencil $\overline M_T$ with
\[
 \overline M_T(x,y)K_T=H_TL(x,y),
 \qquad H_T\in\Mat_2(\Z),\quad \det H_T\ne0.
\]
Thus $K_T$ records the kernel coefficients and $H_T$ records how the
pencil realizes their relations in the output lattice.
These two integral matrices retain the lattice information beyond
the common rational model $L$ (Lemma~\ref{lem:quadratic-kernel-data}).

The advantage of this description is that only the first-factor
matrix $S$ remains to be chosen. Define its symmetric-square action
by $\nu(Sz)=\Sigma_2(S)\nu(z)$. The basic relations transform by
\[
 L(Sz)\Sigma_2(S)=(\det S)S L(z).
\]
Absorbing the primitive-kernel sign into the comparison maps, the
kernel relation therefore forces
$Q=K_{T'}\Sigma_2(S)K_T^{-1}$, and the reduced pencil forces
$P=(\det S)H_{T'}SH_T^{-1}$.
First compare $|\det K_T|$ and $|\det H_T|$ with their target values.
Once they match, the forced maps have determinant of absolute value
one whenever $S\in\GL_2(\Z)$. It remains only to test their
\emph{integrality}:
\[
 H_{T'}SH_T^{-1}\in\Mat_2(\Z),\qquad
 K_{T'}\Sigma_2(S)K_T^{-1}\in\Mat_3(\Z).
\]
Proposition~\ref{prop:symmetric-criterion} proves that these
conditions are equivalent to tensor isomorphism.

\paragraph{Turn integrality into a fixed system of congruences.}
Choose $N=2|\det H_T\det K_T|$ to clear denominators.
The entries of $NH_{T'}SH_T^{-1}$ and
$NK_{T'}\Sigma_2(S)K_T^{-1}$ are integer polynomials in the four
entries of $S$. Requiring them to vanish modulo $N$ gives four
linear and nine quadratic congruences. For each fixed
$\delta\in\{\pm1\}$, add $\det S\equiv\delta\pmod N$.
We obtain fourteen equations of degree at most two in four variables.

This finite test is sufficient because reduction from $\SL_2(\Z)$
to $\SL_2(\Z/N\Z)$ is surjective~\cite{bhj}.
Multiplying by $\diag(\delta,1)$ reduces the determinant-$\delta$
case to this statement. Thus a residue solution lifts to an
integer matrix with determinant \emph{exactly} $\delta$.
The lift preserves the cleared-denominator congruences, hence makes
the two forced maps integral and therefore unimodular.
This is the essential passage from modular feasibility to an integral
tensor isomorphism. The sign $\delta$ is fixed globally, including
across all prime-power factors of $N$
(Theorem~\ref{thm:symmetric-bqp}).

\paragraph{Decide the congruences in polynomial bit complexity.}
The number of equations, variables, and their degrees are fixed, but
the modulus may be exponentially large in its bit length. We use
Theorem~\ref{thm:intro-congruences} to decide solvability without
enumerating residue classes. Given the factorization of $N$, the
Chinese remainder theorem reduces the question to prime powers $p^e$.
Solvability there is expressed by a fixed formula over $\Q_p$, and
the unknowns are integral and each polynomial value has valuation
at least $e$, where the valuation counts powers of $p$.
This retains the full exponent and expresses precisely the finite
congruence condition.

Quantifier elimination turns this fixed formula into a fixed
collection of arithmetic tests. Pas handles all but finitely many
fixed exceptional primes, while elimination in the value group and residue
field leaves integer comparisons and fixed-degree univariate root
tests, while Macintyre handles the exceptional
primes~\cite{pas,Coo72,Kie76,macintyre}.
The resulting tests have polynomial bit complexity
(Lemma~\ref{lem:fixed-format-congruences}). Shor's algorithm supplies the
factorization~\cite{shor}, so this branch is in \BQP.
Only the existence of a residue solution is needed and lifting proves
sufficiency without requiring the congruence algorithm to output one.

\subsubsection{Completing the algorithm}\label{subsec:overview-assembly}

The two routes, together with the elementary cases and the pure-cube
flag test, exhaust all tensors. The structural reductions use
polynomial-time rational and integer linear algebra together with
fixed-degree polynomial calculations, and the lattices, coefficient
matrices, and congruence modulus have
polynomial bit length. The quantum work consists of the ~\cite{bhj}
principality procedures, the known $2\times2\times2$ comparison, and
integer factorization. Combining these procedures with standard
error amplification proves Theorem~\ref{thm:main}.
Algorithm~\ref{alg:integral-233-ti} gives the complete case division.

In the $f_{T}\equiv 0$ case, padding each slice of a $2\times2\times2$ tensor by a zero row and
column preserves and reflects integral isomorphism
(Proposition~\ref{prop:faithful-padding}). Composing this embedding
with the reduction of Grochow--Qiao--Stange--Sun~\cite{gqss} places
their real-quadratic principality problem inside the identically
singular case (Corollary~\ref{cor:principality-obstruction}).
Thus the kernel analysis reaches a part of the tensor problem that
already contains substantial arithmetic structure.

\section{Preliminaries}

\subsection{Tensor and action conventions}
For $T=(A,B)$, write
\begin{equation*}
M_T(x,y)=xA-yB, \qquad f_T(x,y)=\det M_T(x,y).
\end{equation*}
A tensor is \emph{regular} if $f_T\not\equiv0$, and we say it is \emph{nondegenerate} if
$\Disc(f_T)\neq0$ where for a binary cubic $f(x,y)=ax^{3}+bx^{2}y+cxy^{2}+dy^{3}$, we use
\[
\text{Disc}(f)=b^{2}c^{2}-4ac^{3}-4b^{3}d-27a^{2}d^{2}+18abcd.
\]
For nonzero $f$, its discriminant is nonzero if and only if its three projective roots over $\overline{\mathbb{Q}}$ are distinct. For
\(
 h=\begin{pmatrix}\alpha&\beta\\ \gamma&\delta\end{pmatrix}\in\GL_2(\Z)
\)
and $P,Q\in\GL_3(\Z)$, our action is
\begin{equation}
(h,P,Q)\cdot(A,B)=\bigl(P(\alpha A-\beta B)Q^{-1},\,P(-\gamma A+\delta B)Q^{-1}\bigr).
\label{eq:action}
\end{equation}
This is the usual tensor action on the slices $(A,-B)$. Writing the right action as $Q^{-1}$ instead of a
transpose does not change the orbit relation. Define
\begin{equation}
f^h(x,y)=f(\alpha x+\gamma y,\beta x+\delta y),
\qquad
f_{(h,P,Q)\cdot T}=\det(P)\det(Q)^{-1}f_T^h.
\label{eq:form-action}
\end{equation}
We also regard $T$ as the bilinear map
\begin{equation}
\mu_T:E\times F\longrightarrow G,
\qquad
\mu_T((x,y),v)=(xA-yB)v,
\qquad
E=\Z^2,\quad F=G=\Z^3.
\label{eq:bilinear-map}
\end{equation}
Then $T\cong T'$ means that there are unimodular $S,Q,P$ such that
\begin{equation}
P\mu_T(u,v)=\mu_{T'}(Su,Qv).
\label{eq:bilinear-isomorphism}
\end{equation}
The conventions agree exactly: if $T'=(h,P,Q)\cdot T$ in~\eqref{eq:action}, then $S=h^{-\transpose}$ in~\eqref{eq:bilinear-isomorphism}. We use $h$ for
slice transformations and $S$ for first-input maps in bilinear coordinates. In either notation, $Q$ acts
on the second input and $P$ on the output.

For $3\times3$ slices we use
\[
G_3=\{(P,Q)\in\GL_3(\Z)^2:\det P=\det Q\}.
\]
Strict equivalence, denoted $T\sim_{\mathrm{str}}T'$, means equivalence under $(I_2,P,Q)$ with $(P,Q)\in G_3$. It
preserves the determinant form exactly.

\subsection{Lattices and unused directions}
A full rational lattice in $\Q^n$ is a subgroup $B\Z^n$ with $B\in\GL_n(\Q)$. More generally, a lattice may be full in a specified rational subspace wehre its rank is the dimension of that subspace. A sublattice $K\subseteq\Z^n$ is saturated
if $mv\in K$, with $0\neq m\in\Z$ and $v\in\Z^n$, implies $v\in K$.

For a bilinear map $\mu:E\times F\to G$ between free abelian groups, define its unused second-input
directions and its saturated output span by
\begin{equation}
K=\{v\in F:\mu(E,v)=0\},
\qquad
H=\operatorname{span}_{\Q}\mu(E,F)\cap G.
\label{eq:KH}
\end{equation}
The associated reduced map is
\begin{equation}
\bar\mu:E\times(F/K)\longrightarrow H.
\label{eq:reduced-map}
\end{equation}

\section{Basic integral reductions and lattice comparison}

\subsection{Dependent slices}
\begin{lemma}[Dependent slices]\label{lem:dependent-slices}
Isomorphism of integral tensors with linearly dependent slices is
decidable in deterministic polynomial time.
\end{lemma}
\begin{proof}
The zero tensor is recognized directly, so suppose that $(A,B)\neq (0,0)$.
If the two slices are linearly dependent over $\mathbb{Q}$, then the
$\mathbb{Z}$-module generated by their vectorizations has rank one.
Hermite normal form therefore computes a unimodular first-factor change
sending
\[
        (A,B)\longmapsto (A_0,0)
\]
for some nonzero $A_0\in\Mat_3(\mathbb{Z})$.  Thus a tensor with dependent
slices is, after an allowed change of basis, simply one integer matrix
together with an unused first-factor direction.

It remains to compare two tensors $(A_0,0)$ and $(A_0',0)$.  If
\[
h=
\begin{pmatrix}
\alpha & \beta\\
\gamma & \delta
\end{pmatrix}
\in \GL_2(\mathbb{Z}),
\]
then by~\eqref{eq:action},
\[
(h,P,Q)\cdot(A_0,0)
=
\bigl(\alpha P A_0Q^{-1},-\gamma P A_0Q^{-1}\bigr).
\]
Since $A_0\neq 0$, obtaining a zero second slice forces $\gamma=0$.
Unimodularity of $h$ then gives $\alpha=\pm1$.  This sign may be absorbed
into $P$, so
\[
        (A_0,0)\cong(A_0',0)
        \quad\Longleftrightarrow\quad
        A_0'=P A_0Q^{-1}
\]
for some $P,Q\in\GL_3(\mathbb{Z})$.

The latter is exactly unimodular left-right equivalence of integer
matrices, which is decided by Smith normal form.  Hence the dependent-slice
case is decidable in deterministic polynomial time.
\end{proof}

\subsection{Removing unused directions}
The groups $K$ and $H$ from~\eqref{eq:KH} are saturated.

\begin{lemma}\label{lem:remove-unused}
For two maps with the same ambient dimensions and
the same ranks of $K$ and $H$, integral isomorphism is equivalent to integral isomorphism of their
reduced maps~\eqref{eq:reduced-map}.
\end{lemma}
\begin{proof}
An isomorphism preserves the intrinsically defined groups $K$ and $H$ and induces an isomorphism of the reduced maps. Conversely, choose integral bases adapted to these saturated sublattices.
The removed input directions are zero columns and the removed output directions are zero rows.
An isomorphism of the reduced maps extends by any unimodular identifications of the unused
complements.
\end{proof}

\section{Balanced modules for nonzero determinant forms}
This section compares two regular tensors with the same determinant form. The parametrizations of
Bhargava and Wood~\cite{bhargava,wood} allow us to identify the objects to be compared.

\subsection{An invertible first matrix and the associated cubic ring}
Suppose $f_T=f_{T'}=f\not\equiv0$. The polynomial $f(1,z)$ is nonzero and has degree at most three. Hence
some $k\in\{0,1,2,3\}$ satisfies $f(1,k)\neq0$. Make the same replacement on both tensors:
\begin{equation}
(A,B)\longmapsto(A-kB,B).
\label{eq:shear-regular}
\end{equation}
This transformation is induced by
\(
 h=\begin{pmatrix}1&k\\0&1\end{pmatrix},
\)
is reversible, and preserves strict equivalence. After relabeling, we will continue to denote the first matrix in the tuple as $A$ so that $\det A=\det A'\neq0$. This makes explicit the reduction to a nonzero leading coefficient used in Wood's
construction~\cite[Sections~3.1--3.3]{wood}, which produces the associated modules and cubic ring.

Write
\[
f(x,y)=f_0x^3+f_1x^2y+f_2xy^2+f_3y^3,\qquad f_0\neq0,
\]
and set
\[
p_f(z)=z^3+\frac{f_1}{f_0}z^2+\frac{f_2}{f_0}z+\frac{f_3}{f_0}.
\]

Let $\theta$ denote the image of $z$ in
$\mathbb{Q}[z]/(p_f(z))$, so that $p_f(\theta)=0$. The associated rational
algebra and cubic ring are
\[
K_f=\mathbb{Q}[\theta],\qquad
R_f=\mathbb{Z}\cdot 1+\mathbb{Z}\cdot\zeta_1+\mathbb{Z}\cdot\zeta_2,
\qquad
\zeta_1=f_0\theta,\quad
\zeta_2=f_0\theta^2+f_1\theta.
\]
The displayed lattice is a ring, free of rank three over $\Z$, and $R_f\tensor_{\Z}\Q=K_f$~\cite[Section~2]{wood}. The
element $\theta$ need not belong to $R_f$, so its action on an integral module is generally rational rather
than integral. The coefficient map used to recover the
tensor is
\begin{equation}
\pi_f:R_f\longrightarrow\Z^2,
\qquad
r_0+r_1\zeta_1+r_2\zeta_2\longmapsto(r_2,-r_1).
\label{eq:pi-f}
\end{equation}

\begin{definition}[{\cite[Section~3]{wood}}]\label{def:balanced-modules}
A balanced pair for $f$ consists of $R_f$-modules
$M,N$, each free of rank three as an abelian group, and an $R_f$-linear map
\[
\beta:M\tensor_{R_f}N\longrightarrow R_f.
\]
In integral bases, regard $m\in M$ as a row vector and $n\in N$ as a column vector, and write
\[
\pi_f(\beta(m,n))=(mAn,-mBn).
\]
We require the balance condition
\[
    \det(xA-yB)=\pm f(x,y).
\]
A based balanced pair includes choices of bases for which
$\det(xA-yB)=f(x,y)$ exactly.
An isomorphism of balanced pairs consists of two module isomorphisms
preserving $\beta$.
\end{definition}

Here $\beta(m,n)$ abbreviates $\beta(m\tensor n)$; tensoring over $R_f$ imposes
$\beta(rm,n)=\beta(m,rn)$. We write the balanced pairing as $\beta$ to distinguish it from
the vector-valued tensor map $\mu_T$ in~\eqref{eq:bilinear-map}.

\begin{theorem}[\cite{wood}]\label{thm:wood}
For every nonzero integral binary cubic $f$, tensors with determinant form $f$
correspond bijectively to based balanced pairs for $f$. The correspondence is compatible with integral
changes of basis and identifies strict-equivalence classes with isomorphism classes of balanced pairs.
\end{theorem}
\begin{proof}
This is~\cite[Theorem~3.1]{wood}, specialized to degree three and the slices $(A,-B)$. The construction
is explicit in~\cite[Section~3.1]{wood}, and first-factor changes are treated in~\cite[Section~3.3]{wood}.
\end{proof}

When $\Disc(f)\neq0$, the algebra $K_f$ is a product of number fields and the modules can be
represented by fractional ideals. This recovers the balanced-ideal description underlying Bhargava's
theorem~\cite[Theorem~2]{bhargava}; see also~\cite[Proposition~5.4 and Theorem~5.5]{wood}.

\subsection{The compatibility diagram}
Let $L=\operatorname{Hom}_{\Z}(M,\Z)$ with action $(r\psi)(m)=\psi(rm)$. The introduction of $L$ will make our conditions of finding two module isomorphisms that respect the necessary balancing map $\beta$ easier to analyze. Let
$\lambda_f:R_f\to\Z$ read the coefficient of $\zeta_2$. The balanced pairing gives an integral map
\begin{equation}
\phi_T:N\longrightarrow L,
\qquad
\phi_T(n)(m)=\lambda_f(\beta(m,n))=mAn.
\label{eq:phi-T}
\end{equation}
Thus $\phi_T$ is represented by $A$ and is $R_f$-linear by the balancing relation. Let
\begin{equation}
D=A^{-1}B,\qquad E=BA^{-1}.
\label{eq:DE}
\end{equation}
On $N\tensor_{\Z}\Q$ and $L\tensor_{\Z}\Q$, the actions of $\theta$ are respectively $D$ and $E$~\cite[Proposition~3.3]{wood}. In particular,
\begin{equation}
EA=AD=B.
\label{eq:EADB}
\end{equation}

\begin{lemma}\label{lem:compatible-module-isos}
Suppose $T,T'$ have the same nonzero determinant
form and the common replacement~\eqref{eq:shear-regular} has been made. Then $T\sim_{\mathrm{str}}T'$ if and only if there are
integral module isomorphisms $Q:N\to N'$ and $P:L\to L'$ making
\begin{equation}
\begin{tikzcd}[column sep=large,row sep=large,ampersand replacement=\&]
N \arrow[r,"A"] \arrow[d,"Q"'] \& L \arrow[d,"P"]\\
N' \arrow[r,"A'"'] \& L'
\end{tikzcd}
\label{eq:compatibility-diagram}
\end{equation}
commute. Equivalently, there are $P,Q\in\GL_3(\Z)$ satisfying
\begin{equation}
PA=A'Q,\qquad QD=D'Q.
\label{eq:compatibility-equations}
\end{equation}
\end{lemma}
\begin{proof}
A rational map is $R_f$-linear if and only if it commutes with the action of $\theta$ and an integral module
map must additionally preserve the integral lattices. Thus~\eqref{eq:compatibility-equations} expresses the commuting square
and the module condition on $N$. It also implies the module condition on $L$, since
$E=ADA^{-1}$ and $E'=A'D'(A')^{-1}$.

Using $B=AD$, the equations give
\[
PB=PAD=A'QD=A'D'Q=B'Q.
\]
Hence $A'=PAQ^{-1}$ and $B'=PBQ^{-1}$. Since $\det A=\det A'\neq0$, one has $\det P=\det Q$, so this is
strict equivalence. Conversely, a strict equivalence gives the same equations. By Theorem~\ref{thm:wood}, it
preserves the whole balanced pair.
\end{proof}

\subsection{Lattice isomorphism wrapper}
A $\Z$-order in a finite-dimensional rational algebra is a unital subring that is a full lattice. We use
only the principality decision consequence of Bley--Hofmann--Johnston~\cite[Propositions~8.1--8.2 and Remark~8.8(b)]{bhj}.
Given an order and a full left lattice by bases, this decides in quantum polynomial time whether the
lattice is free of rank one over the order, provided the semisimple quotient of the algebra is explicitly
decomposed as a product of full matrix algebras over number fields. No free generator is requested or
used in the surrounding computation. The quotient is by the Jacobson radical, the largest nilpotent
two-sided ideal. We describe the explicit required decomposition below.

\begin{proposition}\label{prop:fixed-form}
Strict equivalence of integral $2\times3\times3$ tensors with
the same nonzero determinant form is decidable in \BQP.
\end{proposition}
\begin{proof}
Make~\eqref{eq:shear-regular} and form $D=A^{-1}B$ and $D'=(A')^{-1}B'$. First test rational similarity using
rational canonical form. If $D,D'$ are not similar, reject by Lemma~\ref{lem:compatible-module-isos}. Otherwise compute
$J\in\GL_3(\Q)$ with $JD=D'J$.

The compatible integral homomorphisms are represented by
\begin{equation}
\mathcal H=\{Q\in\Mat_3(\Z):QD=D'Q,\;A'QA^{-1}\in\Mat_3(\Z)\}.
\label{eq:H-lattice}
\end{equation}
The second map is determined by $P=A'QA^{-1}$. Compatible endomorphisms of the target form the
order
\begin{equation}
\mathcal E=\{X\in\Mat_3(\Z):XD'=D'X,\;A'X(A')^{-1}\in\Mat_3(\Z)\}.
\label{eq:E-order}
\end{equation}
Both lattices are computable by integer linear algebra. For example, keep the integral unknowns
$P,Q$, solve $PA=A'Q$ and $PB=B'Q$, and project to $Q$. Composition makes $\mathcal H$ a left $\mathcal E$-module.

Let
\[
\mathcal C=C_{\Q}(D')=\{X\in\Mat_3(\Q):XD'=D'X\}.
\]
Clearing denominators shows that $\mathcal E$ is a full order in $\mathcal C$. Moreover,
\[
    \mathcal I=\mathcal HJ^{-1}\subseteq\mathcal C
\]
is a full left $\mathcal E$-lattice. Indeed, every rational solution of
$QD=D'Q$ has the form $Q=XJ$ for some $X\in\mathcal C$, and after multiplying
by a suitable nonzero integer it satisfies the integrality conditions
defining $\mathcal H$.

Let $Q_1,\ldots,Q_s$ be a $\Z$-basis of $\mathcal H$, computed by the integer linear algebra above
with polynomial bit length. Define the row lattice
\[
\mathcal R(\mathcal H)=\sum_{i=1}^{s}\Mat_{1\times3}(\Z)Q_{i}\subseteq\Mat_{1\times3}(\Z).
\]
This is the lattice generated by all rows of all matrices in $\mathcal H$. It is computed from the
stacked matrix
\[
M_{\mathcal H}=\begin{pmatrix}Q_1\\ \vdots\\ Q_s\end{pmatrix}.
\]
Since $s\leq9$, this matrix has at most $27$ rows and three columns. Smith normal form decides in
polynomial time whether $\mathcal R(\mathcal H)=\Mat_{1\times3}$. We claim that
\[
T\strictiso T'
\quad\Longleftrightarrow\quad
\mathcal I\text{ is principal over }\mathcal E
\quad\text{and}\quad
\mathcal R(\mathcal H)=\Mat_{1\times3}(\Z).
\]

If $Q_0$ is a compatible isomorphism, every $Q\in\mathcal H$ has $QQ_0^{-1}\in\mathcal E$, so
\[
\mathcal H=\mathcal E Q_0,\qquad \mathcal I=\mathcal E(Q_0J^{-1}).
\]
Moreover, the rows of $Q_0\in\GL_3(\Z)$ generate $\Mat_{1\times3}(\Z)$, so the row-lattice test succeeds.
Conversely, suppose $\mathcal I=\mathcal E\alpha$. Fullness gives $\mathcal C=\mathcal C\alpha$, hence
$\alpha\in\mathcal C^\times$. Set $Q_0=\alpha J$ for the correctness argument; neither $\alpha$ nor
$Q_0$ is computed by the algorithm. Then $\mathcal H=\mathcal E Q_0$. Since every $X\in\mathcal E$
is integral, every row of $XQ_0$ is an integer linear combination of rows of $Q_0$. Conversely,
$I_3\in\mathcal E$ implies $Q_0\in\mathcal H$. Thus
\[
\mathcal R(\mathcal H)=\Mat_{1\times3}(\Z)Q_0.
\]
If the row-lattice test succeeds, $Q_0\in\GL_3(\Z)$. Its companion $P_0=A'Q_0A^{-1}$ is integral
by the definition of $\mathcal H$, and $\det P_0=\det Q_0$ because $\det A=\det A'$. Therefore
$P_0$ is also unimodular, and Lemma~\ref{lem:compatible-module-isos} gives strict equivalence.
This implements the additional isomorphism condition in the compatible-homomorphism reduction
of~\cite[Propositions~3.1 and~8.1]{bhj} using only a principality decision and a polynomial-time
row-lattice computation.

It remains to supply the algebra decomposition. Regard $\mathbb{Q}^3$ as a $\mathbb{Q}[z]$-module with $z$ acting as
$D'$. By the rational canonical form decomposition
\cite[Proposition~9.11]{bhj},
\[
    \mathbb{Q}^3
    \cong
    \bigoplus_{p,j}
    \left(\mathbb{Q}[z]/(p^j)\right)^{m_{p,j}},
\]
where $p$ ranges over monic irreducible polynomials and $m_{p,j}$ is
the multiplicity of the block $\mathbb{Q}[z]/(p^j)$. There is an explicit, efficiently computable
quotient map with
\begin{equation}
\mathcal C/\Jac(\mathcal C)\cong
\prod_{p,j:\,m_{p,j}>0}\Mat_{m_{p,j}}\bigl(\Q[z]/(p)\bigr)
\label{eq:semisimple-quotient}
\end{equation}
by~\cite[Corollary~9.9 and Proposition~9.11]{bhj}. Every $\Q[z]/(p)$ is a number field. The required arithmetic
hypotheses therefore hold, and the cited quantum procedures give the decision algorithm.
\end{proof}

\section{The first-factor action on regular tensors}
The fixed-form test compares the last two factors, so we now account for all first-factor changes that combine the two matrix slices $A$ and $B$.

We may arrange that the last two transformations have equal determinants. Indeed, if
$T'=(h,P,Q)\cdot T$ and $\det P=-\det Q$, replace $(h,P,Q)$ by $(-h,-P,Q)$.
The resulting tensor is unchanged, while $\det(-P)=\det Q$ because the slices have size three.
Thus every full isomorphism is represented with $f_T^h=f_{T'}$, followed by strict equivalence.

\subsection{Finite determinant-form comparison}
A \emph{pure cube} is a form $C\ell^3$ with $C\in\Z, C\neq 0$ and $\ell=ux+vy$ a primitive integral linear form. We note that, by definition, the discriminant of $f_{T}$ is equal to zero only when $f_{T}$ either has a repeated root of multiplicity two or a single repeated root of multiplicity three. In the following lemma, there are essentially two cases we consider, when $\Disc(f_{T}) \not=0$ vs when $\Disc(f_{T}) = 0$ although in the statement of condition $i$ in ~\ref{lem:first-factor}, we combine the repeated root case with the case where $\Disc(f_{T})\not = 0$ since their results are similar.

\begin{lemma}\label{lem:first-factor}
There is a deterministic polynomial-time procedure with the
following properties.
\begin{enumerate}
\item[(i)] For nonzero binary cubics $f,g$ that are not pure cubes, it lists all $h\in\GL_2(\Z)$ satisfying $f^h=g$.
There are at most six such elements.
\item[(ii)] Any pure cube can be transformed into $Cx^{3}$ with $C>0$. Two pure cubes are equivalent precisely when these
positive integers agree. The stabilizer of $Cx^{3}$ consists of
\begin{equation}
h_{t,\delta}=\begin{pmatrix}1&-t\\0&\delta\end{pmatrix},
\qquad t\in\Z,\quad \delta\in\{\pm1\},
\label{eq:pure-cube-stabilizer}
\end{equation}
acting by $(A,B)\mapsto(A+tB,\delta B)$.
\end{enumerate}
\end{lemma}
\begin{proof}
First suppose the discriminant is nonzero. Unequal discriminants rule out integral equivalence.
For equal discriminants, apply Cremona's bilinear-factor construction with source and target arranged so that the rational equivalence matrices $M$ satisfy
\[
g(x,y)=(\det M)^{-1}f((x,y)M)
\]
for the determinant-twisted action~\cite[Proposition~14, Theorem~15, and Section~6]{cremona}.
Retain the integral matrices with determinant $\pm1$, and for each put $s=\det M$ and $h=sM$.
Since $f$ is cubic,
\[
f^h(x,y)=s^3f((x,y)M)=s^4g(x,y)=g(x,y).
\]
The conversion $M\mapsto(\det M)M$ is an involution on $\GL_2(\Z)$, so it gives the complete required list and preserves the bound of at most six elements. The construction factors polynomials of fixed
degree in a fixed number of variables, so it is polynomial time.
\par For discriminant zero, factor over $\Q$ with multiplicities. Primitive integral factors give either
\[
f=C_0\ell^2m
\qquad\text{or}\qquad
f=C_0\ell^3.
\]
where $m$ is also a primitive integral linear form. Bezout's identity allows for the computation of integers $a,b$ such that $ua+vb=1$. This allows us to define a matrix $h'=\begin{pmatrix}
    a & b\\v & -u
\end{pmatrix}$ so that from our definition of how $h$ acts on $\ell$, we get that $\ell^{h'}(x,y)=u(ax+vy)+v(bx-uy)=(ua+vb)x+(uv-vu)y=x$. Thus, this computation extends the coefficients of $\ell$ to a unimodular matrix sending $\ell$ to $x$.

For the remainder of the proof, write a general first-factor transformation as
\[
h=\begin{pmatrix}\alpha&\beta\\ \gamma&\delta\end{pmatrix}
\in \GL_2(\mathbb{Z}),
\]
as in Section~2.1. This is distinct from the particular normalizing matrix
$h'$ constructed above. For a pure cube, a sign change gives $Cx^3$ with $C>0$. If $h$ stabilizes $Cx^3$, then $(\alpha x+\gamma y)^3=x^3$, so $\alpha=1$, $\gamma=0$,
and unimodularity gives $\delta=\pm1$. This proves (ii).

For $C_0\ell^2m$, the same operations give the unique representative
\begin{equation}
Cx^2(rx+sy),\qquad C>0,\quad s>0,\quad \gcd(r,s)=1,\quad 0\le2r\le s.
\label{eq:double-root-normal-form}
\end{equation}
Replacing $y$ by $y+tx$ changes $r$ by $ts$, and changing the sign of $x$ replaces
$r$ by $-r$. We emphasize that we perform this normalization so that we don't have to consider an infinite family of possible binary cubic forms since $r$ can be any integer, and we can instead work with a representative that is equivalent to all possibilities.

The stabilizer of~\eqref{eq:double-root-normal-form} contains the identity and, only if $s\mid2r$, the matrix
\[
\begin{pmatrix}-1&2r/s\\0&1\end{pmatrix}.
\]
so that we maintain integrality. Indeed, preservation of the repeated factor forces $\gamma=0$ and $\alpha=\pm1$ and comparison of the other
factor gives $\delta=1$ and $s\beta=r(1-\alpha)$. Normalize both forms, reject if their representatives differ,
and otherwise compose the recorded normalizations with these stabilizer elements. Different root
multiplicity patterns cannot be equivalent. This proves (i) in the remaining case.
\end{proof}

\begin{corollary}\label{cor:regular-nonpure}
Isomorphism of regular tensors whose determinant form is not a pure cube is in \BQP.
\end{corollary}
\begin{proof}
List the at most six $h$ with $f_T^h=f_{T'}$ and apply Proposition~\ref{prop:fixed-form} to $h\cdot T$ and $T'$ for each. The
determinant-sign adjustment above ensures that every full isomorphism is included.
\end{proof}

\subsection{Pure cubes and the canonical line and plane}
Suppose both forms have been normalized to
\begin{equation}
f_T(x,y)=f_{T'}(x,y)=Cx^3,\qquad C>0.
\label{eq:pure-cube-form}
\end{equation}
Then $\det A=\det A'=C$ and $D=A^{-1}B$ has characteristic polynomial $z^3$, so $D^3=0$ and likewise for
$D'$. Full isomorphism asks for
\begin{equation}
A'=P(A+tB)Q^{-1},\qquad B'=P(\delta B)Q^{-1},\qquad t\in\Z,\quad \delta\in\{\pm1\}.
\label{eq:pure-cube-isomorphism}
\end{equation}

For a nonzero nilpotent $D\in\Mat_{3}(\Q)$ with $D^3=0$, define
\begin{equation}
F_1(D)=\im D\cap\ker D,\qquad F_2(D)=\im D+\ker D.
\label{eq:nilpotent-flags}
\end{equation}

We note that the operator \(D=A^{-1}B\) is the natural object to study because, once \(A\) is invertible, it records the second slice relative to the first since \(B=AD\). Moreover, in the pure-cube case
\[
\det(xA-yB)=Cx^3
\]
implies
\[
\det(xI-yD)=x^3,
\]
so the characteristic polynomial of \(D\) is \(z^3\), and hence \(D^3=0\). Thus the pure-cube condition turns the matrix pencil into the study of a nilpotent operator.

More importantly, the subspaces determined by the kernel and image of \(D\) are preserved by every allowed isomorphism. Indeed, the first-factor stabilizer acts by
\[
(A,B)\longmapsto (A+tB,\delta B),
\qquad t\in\mathbb Z,\quad \delta\in\{\pm1\}.
\]
For the transformed pair,
\[
D_t=(A+tB)^{-1}(\delta B)
    =\delta (I+tD)^{-1}D.
\]
Since \(D^3=0\),
\[
(I+tD)^{-1}=I-tD+t^2D^2,
\]
and therefore
\[
D_t=\delta(D-tD^2).
\]
The factor \(\delta(I+tD)^{-1}\) is invertible and commutes with \(D\), so
\[
\ker D_t=\ker D,\qquad \operatorname{im}D_t=\operatorname{im}D.
\]
After also applying the right basis change \(Q\), the corresponding operator for the target tensor is conjugate to \(D_t\). Hence \(Q\) must carry the canonical subspaces
\[
F_1(D)=\operatorname{im}D\cap\ker D,\qquad
F_2(D)=\operatorname{im}D+\ker D
\]
to the corresponding subspaces for \(D'\). These are therefore intrinsic invariants of the tensor under the entire pure-cube isomorphism problem.

The nested sequence of subspaces
\[
0\subset F_1(D)\subset F_2(D)\subset \mathbb Q^3
\]
is called a \emph{flag}. In our setting \(F_1(D)\) is one-dimensional and \(F_2(D)\) is two-dimensional. A basis \(e_1,e_2,e_3\) is said to be \emph{adapted to this flag} if its initial basis vectors span the successive subspaces of the flag, namely
\[
F_1(D)=\langle e_1\rangle,\qquad
F_2(D)=\langle e_1,e_2\rangle.
\]
We now choose such a basis.

For any subspace \(U\subseteq\mathbb Q^3\), we use the notation
\[
D(U)=\{Dv:v\in U\}
\]
for the image of \(U\) under the linear map \(D\). Thus the relations
\[
D(F_1)=0,\qquad
D(F_2)\subseteq F_1,\qquad
D(\mathbb Q^3)\subseteq F_2
\]
mean concretely that \(D\) sends every vector in \(F_1\) to zero, every vector in \(F_2\) into \(F_1\), and every vector in \(\mathbb Q^3\) into \(F_2\). In the adapted basis this says
\[
De_1=0,\qquad
De_2\in\langle e_1\rangle,\qquad
De_3\in\langle e_1,e_2\rangle.
\]
These conditions force
\[
D=
\begin{pmatrix}
0& *& *\\
0& 0& *\\
0& 0& 0
\end{pmatrix}.
\]
Thus \(D\) is strictly upper triangular in any basis adapted to the flag.

On the output side, let
\[
E=BA^{-1}=ADA^{-1}.
\]
Since \(A\) conjugates \(D\) to \(E\), it carries the flag of \(D\) to the corresponding flag of \(E\):
\[
AF_i(D)=F_i(E).
\]
Choose an output basis \(f_1,f_2,f_3\) adapted to this flag, so that
\[
F_1(E)=\langle f_1\rangle,\qquad
F_2(E)=\langle f_1,f_2\rangle.
\]
Because \(A\) maps each member of the source flag to the corresponding member of the target flag, its matrix in these adapted bases is upper triangular. Finally,
\[
B=AD,
\]
and multiplying an upper-triangular \(A\) by the strictly upper-triangular \(D\) gives
\[
A=
\begin{pmatrix}
a&p&r\\
0&d&q\\
0&0&c
\end{pmatrix},
\qquad
B=
\begin{pmatrix}
0&e&h\\
0&0&g\\
0&0&0
\end{pmatrix}.
\]
Thus the forms in the following Lemma 5.3 are not arbitrary normal forms, they arise directly by choosing integral bases adapted to canonical subspaces that every pure-cube isomorphism must preserve.
\begin{lemma}\label{lem:adapted-form}
The subspaces in~\eqref{eq:nilpotent-flags} have dimensions one and two,
respectively, and satisfy
\[
D(F_1)=0,\qquad D(F_2)\subseteq F_1,\qquad D(\Q^3)\subseteq F_2.
\]
Every pair satisfying~\eqref{eq:pure-cube-form} with $B\neq0$ is strictly equivalent to
\begin{equation}
A=\begin{pmatrix}
a&p&r\\0&d&q\\0&0&c
\end{pmatrix},
\qquad
B=\begin{pmatrix}
0&e&h\\0&0&g\\0&0&0
\end{pmatrix},
\qquad a,d,c>0.
\label{eq:adapted-pair}
\end{equation}
The bases respect the flags of $D=A^{-1}B$ and $E=BA^{-1}$. Every isomorphism~\eqref{eq:pure-cube-isomorphism} preserves these
flags, so $P,Q$ are upper triangular in adapted bases.
\end{lemma}
\begin{proof}
The restriction of $D$ to the nonzero invariant space $\im D$ is nilpotent, so $\im D\cap\ker D\neq0$.
If this intersection had dimension at least two, both $\im D$ and $\ker D$ would have dimension at least
two, contradicting rank--nullity. Thus $\dim F_1=1$, and the dimension formula gives $\dim F_2=2$. The
stated inclusions follow from $D^3=0$ and $D(F_2)=\im D^2\subseteq\im D\cap\ker D$.

Since $E=ADA^{-1}$, the map $A$ sends $F_i(D)$ onto $F_i(E)$. Intersect each flag with its integral
lattice. Saturation allows a basis of the rank-one intersection to extend first to a basis of the
rank-two intersection and then to a basis of the whole lattice. Choose these bases on both sides of
$A$. In these coordinates $A$ is upper triangular, $B=AD$ sends the first subspace to zero, the second
into the first, and the whole space into the second, giving its strictly upper-triangular form. Sign changes make the diagonal entries of $A$ positive. Since the original
matrix satisfies $\det A=C>0$, while a change of bases gives
\[
    \det(PAQ^{-1})=\frac{\det P}{\det Q}\,C,
\]
the resulting positive determinant must still equal $C$. Hence
$\det P=\det Q$, so these basis changes give a strict equivalence.

Replacing $A$ by $A+tB$ changes $D$ to
\begin{equation}
(A+tB)^{-1}B=(I+tD)^{-1}D=D-tD^2.
\label{eq:shear-on-D}
\end{equation}
The factor $I-tD$ is invertible and commutes with $D$, so this map has the same kernel and image as
$D$. Multiplying $B$ by $-1$ also preserves them. The same is true for $E$. Thus every transformation
in~\eqref{eq:pure-cube-isomorphism} carries the source flags to the target flags, which forces upper triangularity.
\end{proof}

\subsection{One integer linear system and one congruence}
Consider two adapted pairs of the form (5.6). An equivalence between them has the form
\[
A'=P(A+tB)Q^{-1},
\qquad
B'=P(\delta B)Q^{-1},
\qquad
t\in\mathbb Z,\quad \delta\in\{\pm1\},
\]
where, by the flag argument of the previous section, \(P,Q\in\mathrm{GL}_3(\mathbb Z)\) may be taken to be upper triangular. Equivalently,
\[
P(A+tB)=A'Q,
\qquad
P(\delta B)=B'Q.
\]

We first account for the possible diagonal signs of \(P\) and \(Q\). Since an upper-triangular unimodular integer matrix has diagonal entries in \(\{\pm1\}\), each diagonal entry of \(P\) and \(Q\) is a sign. Moreover, \(B\) is strictly upper triangular, so \(A+tB\) has the same diagonal entries \(a,d,c\) as \(A\). Comparing diagonal entries in
\[
P(A+tB)=A'Q
\]
therefore gives
\[
P_{11}a=a'Q_{11},\qquad
P_{22}d=d'Q_{22},\qquad
P_{33}c=c'Q_{33}.
\]
Since \(a,d,c,a',d',c'\) are all positive and \(P_{ii},Q_{ii}\in\{\pm1\}\), an equivalence is possible only if
\[
a=a',\qquad d=d',\qquad c=c',
\]
and in that case the corresponding diagonal signs must agree:
\[
P_{ii}=Q_{ii}\qquad (i=1,2,3).
\]

Write these three common signs as
\[
P_{ii}=Q_{ii}=\varepsilon_i,
\qquad
\varepsilon_i\in\{\pm1\},
\]
and package them into the diagonal sign matrix
\begin{equation}
\Sigma=\operatorname{diag}(\varepsilon_1,\varepsilon_2,\varepsilon_3).
\label{eq:pure-diagonal-signs}
\end{equation}
There are only \(2^3=8\) possible choices of \(\Sigma\), and we simply try each one. We must also try the independent sign
\[
\delta\in\{\pm1\}
\]
coming from the pure-cube stabilizer. Thus there are only \(2^4=16\) sign choices in total.

Fix one choice of \(\Sigma\) and \(\delta\). Since \(\Sigma^2=I\), define
\[
P_0:=P\Sigma,\qquad Q_0:=Q\Sigma.
\]
Equivalently,
\[
P=P_0\Sigma,\qquad Q=Q_0\Sigma.
\]
Because the diagonal of both \(P\) and \(Q\) is precisely \((\varepsilon_1,\varepsilon_2,\varepsilon_3)\), multiplying on the right by \(\Sigma\) removes these signs. Hence \(P_0\) and \(Q_0\) are upper \emph{unitriangular}: they are upper triangular and all three of their diagonal entries are \(1\).

Substituting
\[
P=P_0\Sigma,\qquad Q=Q_0\Sigma
\]
into the two equivalence equations and then multiplying both equations on the right by \(\Sigma\) gives
\[
P_0\bigl(\Sigma A\Sigma+t\,\Sigma B\Sigma\bigr)=A'Q_0,
\]
and
\[
P_0\bigl(\delta\Sigma B\Sigma\bigr)=B'Q_0.
\]
Define the sign-adjusted source pair
\[
\widetilde A:=\Sigma A\Sigma,
\qquad
\widetilde B:=\delta\Sigma B\Sigma.
\]
Since \(\delta^2=1\), we have
\[
\Sigma B\Sigma=\delta\widetilde B,
\]
so the first equation becomes
\[
P_0\bigl(\widetilde A+\delta t\,\widetilde B\bigr)=A'Q_0.
\]
Because \(t\) ranges over all integers, so does \(\delta t\). We may therefore rename \(\delta t\) as \(t\). The two equations are then simply
\[
P_0(\widetilde A+t\widetilde B)=A'Q_0,
\qquad
P_0\widetilde B=B'Q_0.
\]

Thus the only purpose of the finite sign enumeration is to remove the possible diagonal signs of \(P\) and \(Q\). For simplicity, for the remainder of this fixed sign choice, relabel
\[
(\widetilde A,\widetilde B)\ \text{as}\ (A,B),
\qquad
(P_0,Q_0)\ \text{as}\ (P,Q).
\]
After this relabeling we may assume that \(P\) and \(Q\) are upper unitriangular and that the equivalence equations are
\[
P(A+tB)=A'Q,
\qquad
PB=B'Q.
\]

Write
\[
A'=
\begin{pmatrix}
a&p'&r'\\
0&d&q'\\
0&0&c
\end{pmatrix},
\qquad
B'=
\begin{pmatrix}
0&e'&h'\\
0&0&g'\\
0&0&0
\end{pmatrix}.
\]
We may now write the remaining unknown upper-unitriangular changes of basis as
\begin{equation}
P=
\begin{pmatrix}
1&u&w\\
0&1&v\\
0&0&1
\end{pmatrix},
\qquad
Q=
\begin{pmatrix}
1&x&z\\
0&1&y\\
0&0&1
\end{pmatrix}.
\label{eq:unitriangular-PQ}
\end{equation}

Before solving for these unknown entries, compare the \((1,2)\) and \((2,3)\) entries of
\[
PB=B'Q.
\]
Because \(P\) and \(Q\) are upper unitriangular, these two entries cannot be changed by the off-diagonal parameters \(u,v,w,x,y,z\). They give the necessary conditions
\[
e=e',
\qquad
g=g'.
\]
If either equality fails, the current sign choice cannot give an equivalence and is rejected. After this check, the remaining equations determine whether suitable integers
\[
u,x,v,y,t,w,z
\]
exist.

The matching entries $e,g$ may vanish.
The equations $P(A+tB)=A'Q$ and $PB=B'Q$ are exactly
\begin{align}
du-ax+et&=p'-p,\label{eq:pure-linear-1}\\
cv-dy+gt&=q'-q,\label{eq:pure-linear-2}\\
gu-ey&=h'-h,\label{eq:pure-linear-3}\\
r'&=r+ht+qu+gut-p'y+cw-az.\label{eq:pure-corner}
\end{align}
All other entries already agree. The first three equations form
\begin{equation}
K\xi=b,
\qquad
K=\begin{pmatrix}
d&-a&0&0&e\\
0&0&c&-d&g\\
g&0&0&-e&0
\end{pmatrix},
\qquad
\xi=\begin{pmatrix}u\\x\\v\\y\\t\end{pmatrix},
\qquad
b=\begin{pmatrix}p'-p\\q'-q\\h'-h\end{pmatrix}.
\label{eq:Kxi}
\end{equation}

\begin{proposition}\label{prop:pure-cube-test}
For two adapted pairs and a fixed sign choice with
matching $a,d,c,e,g$, equivalence is decided by one integer linear system and one congruence.
\end{proposition}
\begin{proof}
Solve~\eqref{eq:Kxi} over $\Z$ and reject this choice if it has no solution. Otherwise choose a solution
$\xi_0=(u_0,x_0,v_0,y_0,t_0)^{\transpose}$ and set
\begin{equation}
r_1=r+ht_0+qu_0+gu_0t_0-p'y_0.
\label{eq:r1}
\end{equation}
Put $\mathcal L=\ker_{\Z}K$ and $\rho=\gcd(a,c)$. For $\xi=(u,x,v,y,t)^{\transpose}\in\mathcal L$, define
\begin{equation}
\Psi(\xi)=h't+q'u-p'y+gut.
\label{eq:Psi}
\end{equation}
For any integral basis $\ell_1,\ldots,\ell_m$ of $\mathcal L$, let
\begin{equation}
\eta=\gcd\bigl(\rho,\Psi(\ell_1),\ldots,\Psi(\ell_m)\bigr).
\label{eq:eta}
\end{equation}
We claim that this sign choice succeeds exactly when
\begin{equation}
r'-r_1\equiv0\pmod\eta.
\label{eq:pure-final-congruence}
\end{equation}

Take $w=z=0$ in the transformation supplied by $\xi_0$. It makes $p,q,h$ equal to $p',q',h'$ and
changes the corner entry to $r_1$. Further transformations fixing these matched entries are precisely
solutions of $K\xi=0$, together with the free $w,z$. Their changes of the corner are
\begin{equation}
\Psi(\xi)+cw-az.
\label{eq:corner-changes}
\end{equation}
Passing to the homogeneous system is legitimate because the upper-triangular transformations and
the integral shears form a group.

The free terms $cw-az$ give exactly $\rho\Z$. The quadratic expression $\Psi$ is additive modulo $\rho$ on
the kernel lattice. To see this, let
$\xi=(u,x,v,y,t)^{\transpose}$ and
$\widetilde\xi=(\widetilde u,\widetilde x,\widetilde v,\widetilde y,\widetilde t)^{\transpose}$ belong to $\mathcal L$. The
homogeneous equations give
\[
ax=du+et,
\qquad
c\widetilde v=d\widetilde y-g\widetilde t,
\qquad
g\widetilde u=e\widetilde y.
\]
Consequently
\begin{equation}
\begin{aligned}
\Psi(\xi+\widetilde\xi)-\Psi(\xi)-\Psi(\widetilde\xi)
&=g(u\widetilde t+\widetilde u t)\\
&=ax\widetilde y-cu\widetilde v\equiv0\pmod\rho.
\end{aligned}
\label{eq:Psi-additive}
\end{equation}
Thus $\Psi\bmod\rho:\mathcal L\to\Z/\rho\Z$ is a group homomorphism. Its image is generated by the $\Psi(\ell_i)$, and its
preimage in $\Z$ is $\eta\Z$.

This describes exactly the attainable translations. Every value in~\eqref{eq:corner-changes} is divisible by $\eta$.
Conversely, choose a lattice vector attaining the residue of any multiple of $\eta$ modulo $\rho$, and use
$cw-az$ to supply the remaining multiple of $\rho$. The set of possible corner changes is therefore $\eta\Z$,
proving~\eqref{eq:pure-final-congruence}.
\end{proof}

\begin{corollary}\label{cor:pure-cube}
Isomorphism on the regular pure-cube family is decidable in deterministic polynomial time.
\end{corollary}
\begin{proof}
First compare $\dim_\Q\operatorname{span}\{A,B\}$ and $\dim_\Q\operatorname{span}\{A',B'\}$, and reject if they differ.
If both dimensions are one, apply Lemma~\ref{lem:dependent-slices}. In the remaining regular case both pairs of slices are independent.
Normalize both forms to $Cx^3$ and reject if their positive coefficients differ. The normalized second slices $B,B'$ are nonzero, so Lemma~\ref{lem:adapted-form} applies. Compute the adapted pairs, compare $(a,d,c)$, and
enumerate the signs needed to make $P$ and $Q$ unitriangular. For each choice compare $e,g$ and apply Proposition~\ref{prop:pure-cube-test}. The flag bases, integer
linear solutions, kernel bases, and gcds are efficiently computable. No division by $e$ or $g$ occurs, so
the same calculation applies when either vanishes.
\end{proof}

\begin{corollary}[The regular locus]\label{cor:regular-locus}
Integral $2\times3\times3$ tensor isomorphism restricted to $f_T\not\equiv0$ is in \BQP.
\end{corollary}

\begin{proof}
Being a pure cube is invariant under integral changes of variables. Reject if exactly one form
is a pure cube. Otherwise apply Corollary~\ref{cor:regular-nonpure} or Corollary~\ref{cor:pure-cube}. These alternatives exhaust the
nonzero binary cubics.
\end{proof}

\section{The zero-determinant case}\label{sec:zero-boundary}
We now assume $f_T\equiv0$. The determinant no longer provides a nonzero cubic ring parameter.
After the elementary reductions below, we instead use a polynomial vector in the kernel of
$xA-yB$. Its least possible degree is either one or two. A linear kernel vector gives a
compatible linear diagram, while a quadratic kernel vector gives two coefficient matrices
whose comparison reduces to a fixed system of congruences.

\subsection{Normal rank at most one}
The normal rank is the rank of $xA-yB$ over $\Q(x,y)$. On the zero-form locus it is at most two and
equals
\begin{equation}
\max\{\rank A,\rank(A-B),\rank(A-2B)\}.
\label{eq:normal-rank}
\end{equation}
Indeed, a nonzero maximal minor of $A-tB$ has degree at most two and cannot vanish at all three
values $t=0,1,2$.

\begin{lemma}\label{lem:low-normal-rank}
Isomorphism on the normal-rank-at-most-one locus is decidable
in deterministic polynomial time.
\end{lemma}
\begin{proof}
By Lemma~3.1, we may assume that the two slices are linearly
independent. Since the normal rank is at most one, every specialization
$xA-yB$ has rank at most one. In particular, $A$ and $B$ are nonzero
rank-one matrices, so over $\mathbb{Q}$ we may write
\[
        A=uv^{T},
        \qquad
        B=st^{T}
\]
with nonzero $u,s,v,t\in\mathbb{Q}^{3}$.

We first record the elementary structure forced by the normal-rank
assumption. Suppose that both pairs $u,s$ and $v,t$ were linearly
independent. After rational changes of basis in the output and input
spaces, we could arrange that $u,s$ are the first two output basis
vectors and that $v^{T},t^{T}$ are the first two coordinate
covectors. In these bases, $A+B$ has a nonsingular $2\times2$
diagonal block and hence has rank two. This contradicts the assumption
that every linear combination of $A$ and $B$ has rank at most one.
Therefore either $u,s$ are dependent or $v,t$ are dependent. These
two alternatives cannot both occur, since then $B$ would be a rational
scalar multiple of $A$, contrary to the independence of the slices.

Suppose first that $u,s$ are dependent. Then all images of the two
slices lie in the same rational line
\[
        \operatorname{im}A+\operatorname{im}B.
\]
Let $h$ be a primitive generator of the saturation of this line in
$\mathbb{Z}^{3}$. We may then write
\[
        A=h a^{T},
        \qquad
        B=h b^{T}
\]
with $a,b\in\mathbb{Z}^{3}$. Lemma~3.2 deletes the two unused output
directions, so the tensor is represented exactly by the integer matrix
\[
        \begin{pmatrix}
        a^{T}\\
        b^{T}
        \end{pmatrix}
        \in\operatorname{Mat}_{2\times3}(\mathbb{Z}).
\]
Tensor isomorphism is therefore ordinary unimodular left-right
equivalence of this matrix, which is decided by Smith normal form.

Suppose instead that $v,t$ are dependent. Then the two slices have a
common two-dimensional rational kernel
\[
        K=\ker A\cap\ker B.
\]
After intersecting with $\mathbb{Z}^{3}$ and saturating, Lemma~3.2
deletes these two unused second-input directions. The quotient has rank
one, so the reduced tensor is a $2\times1\times3$ tensor, equivalently
a $3\times2$ integer matrix whose two columns are the images of a
generator of the remaining input direction under the two slices.
Again, tensor isomorphism is unimodular left-right matrix equivalence
and is decided by Smith normal form.

Finally, the two branches cannot be confused by an isomorphism. In the
first case the total image has dimension one and the total row space
has dimension two, while in the second case these dimensions are two
and one, respectively. These dimensions are invariant under the tensor
action. All required lattice saturations, basis changes, and Smith
normal forms are computable in deterministic polynomial time.
\end{proof}

\subsection{Two nonzero kernels}
Assume from now on that the slices are independent and have normal rank two. Compute over $\Q$
\begin{equation}
K_R=\ker A\cap\ker B,
\qquad
K_L=\ker A^{\transpose}\cap\ker B^{\transpose}.
\label{eq:common-kernels}
\end{equation}
Each dimension is zero or one and we regard $K_L$ as a subspace of the dual output space.

If both kernels are nonzero, then each is a one-dimensional rational
subspace. Choose primitive integral generators for
\[
        K_R\cap\mathbb{Z}^{3}
        \qquad\text{and}\qquad
        K_L\cap\mathbb{Z}^{3},
\]
and extend them to unimodular integral bases. On the input side, place
the generator of $K_R$ in the third coordinate. Since
\[
        K_R=\ker A\cap\ker B,
\]
both slices vanish on this basis vector, so the third column of both
$A$ and $B$ is zero.

Similarly, a generator $w$ of $K_L$ satisfies
\[
        w^{T}A=w^{T}B=0.
\]
Choose the output basis so that $w^{T}$ is the third coordinate
covector. Then the third row of both slices is zero. Thus, in these
integral bases,
\[
        xA-yB=
        \begin{pmatrix}
            C(x,y) & 0\\
            0      & 0
        \end{pmatrix},
        \qquad
        C(x,y)\in\Mat_{2}(\mathbb{Z}[x,y]).
\]
In other words, the common right kernel removes one input direction
and the common left kernel removes one output direction, leaving a
$2\times2$ matrix pencil.

The core $C(x,y)$ still has normal rank two. Indeed, deleting a common
zero row and column does not change the rank of the pencil, and the
original pencil has normal rank two by assumption. By Lemma~3.2, the
original tensors are therefore isomorphic exactly when these reduced
$2\times2$ cores are isomorphic under
$\GL_{2}(\mathbb{Z})^{3}$. The \BQP algorithm of
Grochow--Qiao--Stange--Sun consequently handles this branch
\cite[Theorem~3.7]{gqss}. The core is thus a smaller tensor
isomorphism problem, rather than a single fixed rational orbit.

\subsection{A polynomial kernel of degree one or two}
First compare the ordered dimensions of $K_R,K_L$ for the two tensors, rejecting if they differ.
The two-common-kernel case has just been handled. If the remaining pattern has $K_R\neq0$,
transpose both pencils. Simultaneous transposition preserves the isomorphism question so we may therefore assume
throughout the rest of the singular comparison that
\[
K_R=0.
\]
We do not yet delete an output direction: the linear-kernel argument will work whether or not
$K_L$ is nonzero.

A homogeneous polynomial kernel vector of degree $d$ has the form
\begin{equation}
v_T(x,y)=\sum_{j=0}^{d}x^{d-j}y^jv_j,
\qquad v_j\in\Q^3,
\qquad M_T(x,y)v_T(x,y)=0.
\label{eq:polynomial-kernel}
\end{equation}
Comparing coefficients shows that its vectors are exactly the solutions of the rational linear system
\begin{equation}
Av_0=0,\qquad Av_j=Bv_{j-1}\quad(1\le j\le d),\qquad Bv_d=0.
\label{eq:kernel-coefficient-system}
\end{equation}
We therefore first solve this system with $d=1$, and use
$d=2$ only if the first system has no nonzero solution.

\begin{lemma}\label{lem:polynomial-kernel}
Suppose $M_T$ has normal rank two and $K_R=0$. The least degree $d_T$ of a nonzero
homogeneous polynomial kernel vector is either one or two. At that degree, the rational
solution space of~\eqref{eq:kernel-coefficient-system} is one-dimensional. It has a primitive
integral generator, unique up to sign, whose coefficient vectors are computable in deterministic
polynomial time. The degree $d_T$ is invariant under tensor isomorphism.
\end{lemma}
\begin{proof}
Since $M_T(x,y)$ has normal rank two, we may choose two of its rows
that are linearly independent over $\mathbb{Q}(x,y)$. Their cross
product is nonzero and is orthogonal to both rows. The third row lies
in their $\mathbb{Q}(x,y)$-span, since the whole pencil has rank two
over $\mathbb{Q}(x,y)$, and hence it is orthogonal to the same vector.
Therefore the cross product gives a nonzero polynomial vector in the
right kernel of $M_T(x,y)$.

Each entry of $M_T(x,y)$ is a homogeneous linear form in $x,y$, so the
coordinates of this cross product are $2\times2$ minors and are
homogeneous polynomials of degree two. If these three coordinates have
a nonconstant common polynomial factor, divide it out. This does not
change the kernel relation, because if
\[
        M_T(x,y)\bigl(g(x,y)v(x,y)\bigr)=0
\]
and $g$ is nonzero, then
\[
        M_T(x,y)v(x,y)=0
\]
over the integral domain $\mathbb{Q}[x,y]$. We therefore obtain a
nonzero homogeneous polynomial kernel vector whose coordinates have no
common nonconstant factor and whose degree is at most two.

This vector cannot have degree zero. A degree-zero kernel vector would
be a nonzero constant vector $v\in\mathbb{Q}^{3}$ satisfying
\[
        (xA-yB)v=0
\]
identically in $x,y$. Comparing the coefficients of $x$ and $y$ would
give
\[
        Av=Bv=0,
\]
so $v\in K_R$, contradicting the assumption $K_R=0$. Hence the least
possible degree $d_T$ is either one or two.

It remains to show uniqueness at the least degree. Since $M_T$ has
normal rank two, its kernel over the field $\mathbb{Q}(x,y)$ is
one-dimensional. Let
\[
        v(x,y)\in\mathbb{Q}[x,y]^3
\]
be a polynomial kernel vector of least degree whose coordinates have
no common nonconstant factor. Any other polynomial kernel vector
$w(x,y)$ lies in the same one-dimensional kernel over
$\mathbb{Q}(x,y)$, so
\[
        w=\frac{a}{b}v
\]
for coprime $a,b\in\mathbb{Q}[x,y]$. Since every coordinate of $w$ is
a polynomial, $b$ divides every coordinate of $a v$. Coprimality of
$a$ and $b$ then forces $b$ to divide every coordinate of $v$. By our
choice of $v$, its coordinates have no common nonconstant factor, so
$b$ must be constant. Thus every polynomial kernel vector is a
polynomial multiple of $v$.

If $w$ has the same least degree as $v$, that polynomial multiplier
must have degree zero. Hence $w$ is a rational scalar multiple of
$v$. Therefore the solution space of the coefficient equations
\[
        Av_0=0,\qquad
        Av_j=Bv_{j-1}\ (1\leq j\leq d_T),\qquad
        Bv_{d_T}=0
\]
is one-dimensional over $\mathbb{Q}$ at the least degree $d_T$.

Choose any nonzero rational solution of this system. Clearing
denominators gives an integral coefficient vector, and dividing all of
its integer coefficients by their gcd gives a primitive integral
generator. Since the rational solution space is one-dimensional, any
two primitive integral generators differ only by sign.

Finally, $d_T$ is invariant under tensor isomorphism. A change of basis
in the second input multiplies a kernel vector by an invertible
constant matrix, while a change of basis in the first factor makes an
invertible linear substitution in the variables $x,y$. Both operations
send homogeneous polynomial kernel vectors to homogeneous polynomial
kernel vectors of the same degree, and both are reversible. Hence the
least possible kernel degree is preserved.

To compute $d_T$, solve the rational linear system above first for
$d=1$ and, if it has no nonzero solution, for $d=2$. These systems have
fixed size. Rational linear algebra, clearing denominators, and taking
gcds all have polynomial bit complexity, so the primitive generator is
computable in deterministic polynomial time.
\end{proof}

The remaining comparison now has just two cases, distinguished by this one kernel computation.
Section~\ref{sec:kernel-comparison} treats both. The degree one case reduces directly to the
Bley-Hofmann-Johnston algorithms, and degree two case reduces to the direct congruence test
proved in Section~\ref{sec:symmetric-comparison}.
\section{Comparison through the polynomial kernel}\label{sec:kernel-comparison}
Assume the preliminary reductions of Section~\ref{sec:zero-boundary} have been made, so both
pencils have normal rank two and $K_R=0$. Compute their least kernel degrees by
Lemma~\ref{lem:polynomial-kernel} and reject if they differ.

\subsection{A linear kernel gives a compatible diagram}
Suppose $v_T(x,y)=xv_0+yv_1$. Set
\[
\mathcal U_T=\operatorname{span}_{\Q}\{v_0,v_1\},\qquad
\mathcal W_T=\Q Av_1.
\]
The coefficient equations give $Av_0=0$, $Av_1=Bv_0$, and $Bv_1=0$.
The vectors $v_0,v_1$ are independent, since otherwise canceling a common linear polynomial
would give a constant kernel vector. Also $Av_1\neq0$, since its vanishing would put both
coefficient vectors in $K_R$. Thus $\mathcal U_T$ is a plane, $\mathcal W_T$ is a line, and
\[
A\mathcal U_T+B\mathcal U_T=\mathcal W_T.
\]
These subspaces are intrinsic and the uniqueness of the least-degree kernel preserves its coefficient
span, and the displayed image span then preserves the output line. Intuitively, both slices
send one canonical input plane into one canonical output line.

Choose integral bases adapted to $\mathcal U_T\cap\Z^3$ and $\mathcal W_T\cap\Z^3$.
In these adapted bases, write
\[
U_T=\langle e_2,e_3\rangle_{\mathbb Q},
\qquad
W_T=\langle f_1\rangle_{\mathbb Q}.
\]
The kernel relations
\[
Av_0=0,\qquad Av_1=Bv_0,\qquad Bv_1=0
\]
imply
\[
A(U_T)=B(U_T)=W_T.
\]
Indeed,
\[
A(U_T)=\langle Av_0,Av_1\rangle
      =\langle Av_1\rangle=W_T,
\]
and similarly
\[
B(U_T)=\langle Bv_0,Bv_1\rangle
      =\langle Av_1\rangle=W_T.
\]
Thus the last two input basis vectors \(e_2,e_3\) are sent by both \(A\) and \(B\) into the first output coordinate line. Consequently, the second and third columns of both matrices have zeros in their last two entries, so
\[
A=
\begin{pmatrix}
* & * & *\\
* & 0 & 0\\
* & 0 & 0
\end{pmatrix},
\qquad
B=
\begin{pmatrix}
* & * & *\\
* & 0 & 0\\
* & 0 & 0
\end{pmatrix}.
\]
Therefore, with $z=(x,y)^{\transpose}$,
the pencil has the form
\begin{equation}
M_T(z)=\begin{pmatrix}
\ell_Tz&(R_Tz)^{\transpose}\\
C_Tz&0_{2\times2}
\end{pmatrix},
\qquad
\ell_T\in\Mat_{1\times2}(\Z),\quad C_T,R_T\in\Mat_2(\Z).
\label{eq:linear-kernel-block}
\end{equation}
Here $R_T$ is nonsingular, since a nonzero vector in $\ker R_T^{\transpose}$ would give a
common right kernel vector. Also $C_T\neq0$, since the normal rank is two.

The form in~(7.1) is simply the original pencil \(M_T(x,y)=xA-yB\), written in coordinates adapted to the canonical subspaces \(U_T\) and \(W_T\).  Putting \(U_T\) in the last two input coordinates means choosing an input basis for which
\[
U_T=\langle e_2,e_3\rangle_{\mathbb Q},
\]
while putting \(W_T\) in the first output coordinate means choosing an output basis for which
\[
W_T=\langle f_1\rangle_{\mathbb Q}.
\]
Since \(AU_T+BU_T=W_T\), the last two columns of both slices have entries only in the first output coordinate.  Thus, writing \(z=(x,y)^T\), the five potentially nonzero entries of \(M_T(z)=xA-yB\) are naturally grouped as the scalar linear form \(\ell_Tz\), the two entries \(C_Tz\) in the lower-left block, and the two entries \(R_Tz\) in the upper-right block.

Consider the map
\[
X_T=\begin{pmatrix}\ell_T\\ C_T\\ R_T\end{pmatrix}:\mathbb Z^2\longrightarrow\mathbb Z^5
\]
which simply records these five linear forms at once.  The decomposition
\[
\mathbb Z^5=\mathbb Z\oplus\mathbb Z^2\oplus\mathbb Z^2
\]
corresponds respectively to the \(\ell_T\), \(C_T\), and \(R_T\) coordinates.  We retain
\[
\Lambda_1=\mathbb Z\oplus\mathbb Z^2\oplus0,
\qquad
\Lambda_2=\mathbb Z\oplus0\oplus\mathbb Z^2
\]
so that an isomorphism of the flattened maps \(X_T\) must respect this block structure rather than mixing the \(C_T\) and \(R_T\) coordinates arbitrarily.  Thus \(X_T\), together with \(\Lambda_1,\Lambda_2\), contains exactly the structured data of the pencil in~(7.1). 
We compare $X_T$ together with the inclusions of these two sublattices.

\begin{lemma}\label{lem:linear-kernel-diagram}
Two pencils in~\eqref{eq:linear-kernel-block} are integrally tensor-isomorphic if and only if
there are $S\in\GL_2(\Z)$ and $F\in\GL_5(\Z)$ such that
\begin{equation}
FX_T=X_{T'}S,\qquad F\Lambda_1=\Lambda_1,\qquad F\Lambda_2=\Lambda_2.
\label{eq:linear-kernel-compatibility}
\end{equation}
Equivalently, the first condition says that the following
diagram commutes:
\[
\begin{tikzcd}[column sep=large,row sep=large]
\mathbb Z^2
    \arrow[r,"X_T"]
    \arrow[d,"S"']
&
\mathbb Z^5
    \arrow[d,"F"]
\\
\mathbb Z^2
    \arrow[r,"X_{T'}"']
&
\mathbb Z^5 .
\end{tikzcd}
\]
The additional requirements
\[
F\Lambda_1=\Lambda_1,
\qquad
F\Lambda_2=\Lambda_2
\]
ensure that the right-hand change of basis preserves the decomposition of
the five coordinates into the \(\ell_T\), \(C_T\), and \(R_T\) parts,
rather than allowing an arbitrary change of basis in \(\mathbb Z^5\).
\end{lemma}
\begin{proof}
The two distinguished sublattices force the required block form of \(F\).
Recall that
\[
\mathbb Z^5
=
\mathbb Z\oplus\mathbb Z^2\oplus\mathbb Z^2,
\qquad
\Lambda_1=\mathbb Z\oplus\mathbb Z^2\oplus 0,
\qquad
\Lambda_2=\mathbb Z\oplus 0\oplus\mathbb Z^2.
\]
Their intersection is
\[
\Lambda_1\cap\Lambda_2=\mathbb Z\oplus 0\oplus 0.
\]
Hence, if \(F\Lambda_i=\Lambda_i\) for \(i=1,2\), then \(F\) must preserve
this common rank-one sublattice.  Its action there is therefore multiplication
by some \(\varepsilon\in\{\pm1\}\).

Next, preservation of \(\Lambda_1\) implies that a vector in the middle
\(\mathbb Z^2\)-block cannot acquire a component in the last
\(\mathbb Z^2\)-block, while preservation of \(\Lambda_2\) implies that a
vector in the last block cannot acquire a component in the middle block.
Components from either block may, however, be added to the common first
coordinate.  Thus \(F\) necessarily has the form
\[
F=
\begin{pmatrix}
\varepsilon & a & b\\
0 & U & 0\\
0 & 0 & V
\end{pmatrix},
\qquad
a,b\in\Mat_{1\times2}(\mathbb Z),
\qquad
U,V\in\Mat_{2}(\mathbb Z).
\]
Finally, because \(F\) preserves \(\Lambda_1\) and \(\Lambda_2\) as lattices
with equality, its induced maps on
\[
\Lambda_1/(\Lambda_1\cap\Lambda_2)\cong\mathbb Z^2,
\qquad
\Lambda_2/(\Lambda_1\cap\Lambda_2)\cong\mathbb Z^2
\]
must be unimodular. These induced maps are precisely the blocks $U$
and $V$. Hence
\[
U,V\in\GL_2(\mathbb Z).
\]
Conversely, every matrix $F$ of this form preserves both
$\Lambda_1$ and $\Lambda_2$. We now relate such an $F$ to the
usual tensor change-of-basis matrices $P$ and $Q$ from
\[
        P M_T(z)=M_{T'}(Sz)Q.
\]
Recall that the canonical input plane has been placed in the last two
coordinates and the canonical output line in the first coordinate.
Hence every tensor isomorphism has block forms
\[
        P=
        \begin{pmatrix}
            \varepsilon & a\\
            0           & U
        \end{pmatrix},
        \qquad
        Q=
        \begin{pmatrix}
            q & 0\\
            c & W
        \end{pmatrix},
\]
where
\[
        \varepsilon,q\in\{\pm1\},\qquad
        a\in\Mat_{1\times2}(\mathbb Z),\qquad
        c\in\Mat_{2\times1}(\mathbb Z),
        \qquad
        U,W\in\GL_2(\mathbb Z).
\]
Indeed, $P$ must preserve the distinguished output line, while $Q$
must preserve the distinguished input plane. Replacing $P$ and $Q$
simultaneously by their negatives if necessary, we may assume $q=1$.

It is convenient to parametrize the remaining blocks of $Q$ using the
same $b$ and $V$ that occur in $F$. Since $V\in\GL_2(\mathbb Z)$, put
\[
        W=\varepsilon V^{-T},
        \qquad
        c=-V^{-T}b^T,
        \qquad
        V^{-T}:=(V^{-1})^T.
\]
The reason for this choice is the upper-right block of the pencil, which is
written as the row vector $(R_Tz)^T$.  Indeed, comparing the upper-right
blocks in
\[
        P M_T(z)=M_{T'}(Sz)Q
\]
gives
\[
        \varepsilon (R_Tz)^T=(R_{T'}Sz)^T W.
\]
After transposing, this becomes
\[
        \varepsilon R_Tz=W^T R_{T'}Sz.
\]
Thus setting $W=\varepsilon V^{-T}$ makes this equation exactly
\[
        V R_T=R_{T'}S,
\]
which is the $R_T$-block equation occurring in $F X_T=X_{T'}S$.
With this choice of $V$, the upper-left block contains the term
$(R_{T'}Sz)^Tc$.  Setting
\[
        c=-V^{-T}b^T
\]
makes
\[
        (R_{T'}Sz)^Tc=-bR_Tz,
\]
so that moving this term to the other side produces the required
$+\,bR_T$ contribution in the first block equation.  Hence the inverse
transpose and the formula for $c$ are forced by expressing the usual tensor
isomorphism equations in the block coordinates used by $F$.
This is only a reparametrization of the arbitrary unimodular block $W$
and column $c$. Thus we may write
\[
        P=
        \begin{pmatrix}
            \varepsilon & a\\
            0           & U
        \end{pmatrix},
        \qquad
        Q=
        \begin{pmatrix}
            1 & 0\\
            -V^{-T}b^T & \varepsilon V^{-T}
        \end{pmatrix}.
\]
The inverse transpose appears because the $R_T$ block of the pencil is
the row vector $(R_Tz)^T$. With these choices, direct multiplication in $PM_T(z)=M_{T'}(Sz)Q$ now gives
\begin{equation}
UC_T=C_{T'}S,\qquad VR_T=R_{T'}S,\qquad
\varepsilon\ell_T+aC_T+bR_T=\ell_{T'}S.
\label{eq:linear-kernel-block-equations}
\end{equation}
These are $FX_T=X_{T'}S$. Conversely, the displayed formulas recover unimodular $P,Q$ from
any solution of~\eqref{eq:linear-kernel-compatibility}.
\end{proof}

\begin{proposition}\label{prop:linear-kernel-bqp}
Integral isomorphism in the linear-kernel case is decidable in \BQP.
\end{proposition}
\begin{proof}
By Lemma~7.1, the tensors are isomorphic exactly when there are
\[
        S\in\GL_2(\mathbb Z),
        \qquad
        F\in\GL_5(\mathbb Z)
\]
such that
\[
        FX_T=X_{T'}S,
        \qquad
        F\Lambda_1=\Lambda_1,
        \qquad
        F\Lambda_2=\Lambda_2.
\]
We first forget the requirement that $S$ and $F$ be invertible and
consider all compatible integral homomorphisms. Define
\[
H=
\left\{
(S,F)\in
\Mat_2(\mathbb Z)\times\Mat_5(\mathbb Z)
\;\middle|\;
FX_T=X_{T'}S,\;
F\Lambda_i\subseteq\Lambda_i\ (i=1,2)
\right\}.
\]
The conditions defining $H$ are linear in the entries of $S$ and $F$.
Indeed, the equation $FX_T=X_{T'}S$ is linear, while preservation of
the fixed sublattices $\Lambda_1,\Lambda_2$ simply forces the
corresponding off-block entries of $F$ to vanish. Thus $H$ is a
$\mathbb Z$-lattice computable by ordinary integer linear algebra.

Similarly, using the target diagram on both sides, define its
endomorphism order
\[
E=
\left\{
(S,F)\in
\Mat_2(\mathbb Z)\times\Mat_5(\mathbb Z)
\;\middle|\;
FX_{T'}=X_{T'}S,\;
F\Lambda_i\subseteq\Lambda_i\ (i=1,2)
\right\}.
\]
Composition is componentwise multiplication,
\[
        (S_1,F_1)(S_2,F_2)
        =(S_1S_2,F_1F_2),
\]
so $E$ is a ring and acts on $H$ on the left by composition.

The reason for introducing $H$ and $E$ is that an actual isomorphism
would generate all compatible homomorphisms. Suppose, for example,
that
\[
        q=(S_0,F_0)\in H
\]
is an isomorphism, so both $S_0$ and $F_0$ are unimodular. For any
$h\in H$, the composition $hq^{-1}$ is an integral endomorphism of
the target diagram and therefore belongs to $E$. Hence
\[
        H=Eq.
\]
Thus the existence of an isomorphism turns the homomorphism lattice
into a free rank-one left $E$-module. The remaining problem is to
decide this principality condition and then check that a generator is
actually unimodular.

We next describe the rational algebra in which this principality
problem lives. Over $\mathbb Q$, every diagram in the linear-kernel
case can be put into a simple normal form. Since $R_T$ is nonsingular,
rational changes of basis first make
\[
        R_T=I_2.
\]
Once $R_T=I_2$, the free $b$-block in the matrix $F$ from Lemma~7.1
can be chosen to eliminate $\ell_T$, so that
\[
        \ell_T=0.
\]
Finally, ordinary rational row and column operations on $C_T$, while
using the same source change on the $R_T$-block, put
\[
        C_T=C_0:=\diag(I_r,0),
        \qquad
        r=\rank C_T\in\{1,2\}.
\]
Consequently two such diagrams can be rationally isomorphic only if
their values of $r$ agree. If the ranks are equal, these reductions
explicitly give a rational diagram isomorphism
\[
        j:T_{\mathbb Q}\longrightarrow T'_{\mathbb Q}.
\]

We now compute the rational endomorphism algebra of the normalized
target diagram. Write this algebra as $\mathcal A$. In the coordinates
\[
        R_{T'}=I_2,
        \qquad
        \ell_{T'}=0,
        \qquad
        C_{T'}=C_0,
\]
Lemma~7.1 shows that a rational compatible map has
\[
F=
\begin{pmatrix}
e & a & b\\
0 & U & 0\\
0 & 0 & V
\end{pmatrix}.
\]
The three block equations from Lemma~7.1 now become
\[
        V=S,
        \qquad
        UC_0=C_0S,
        \qquad
        aC_0+b=0.
\]
Hence every element of $\mathcal A$ has the form
\[
\left(
S,\,
\begin{pmatrix}
e & a & -aC_0\\
0 & U & 0\\
0 & 0 & S
\end{pmatrix}
\right),
\qquad
UC_0=C_0S.
\]

The semisimple quotient of this algebra is also explicit. An
endomorphism induces independently its scalar action $e$, its action
on $\im C_0$, its action on $\coker C_0$, and its action on
$\ker C_0$. This gives a surjective algebra homomorphism
\[
\mathcal A
\longrightarrow
\mathbb Q
\times
\Mat_r(\mathbb Q)
\times
\Mat_{2-r}(\mathbb Q)
\times
\Mat_{2-r}(\mathbb Q),
\]
where zero-size factors are omitted. The kernel consists only of the
off-diagonal parts that act trivially on these induced spaces. From
the displayed block form, a product of three elements of this kernel
is zero. Thus the kernel is nilpotent. Since the quotient is
semisimple, this kernel is precisely the Jacobson radical, and hence
\[
\mathcal A/\Jac(\mathcal A)
\cong
\mathbb Q
\times
\Mat_r(\mathbb Q)
\times
\Mat_{2-r}(\mathbb Q)
\times
\Mat_{2-r}(\mathbb Q).
\]
In particular, the semisimple quotient is a product of full matrix
algebras over $\mathbb Q$, exactly the form required by the
Bley--Hofmann--Johnston algorithms.

It remains to place the homomorphism lattice inside this single
endomorphism algebra. Since $j$ is a rational isomorphism from the
source diagram to the target diagram, define
\[
        I:=Hj^{-1}.
\]
For $h\in H$, the element $hj^{-1}$ is a rational endomorphism of the
target diagram, so
\[
        I\subseteq\mathcal A.
\]
Moreover, $I$ is stable under left multiplication by $E$, and after
tensoring with $\mathbb Q$ it spans $\mathcal A$. Thus $I$ is a full
left $E$-lattice in $\mathcal A$.

We can therefore apply the principality and lattice-isomorphism
algorithms of
\cite[Propositions~3.1 and~8.1--8.2, Remark~8.8]{bhj}.
They decide in \BQP whether $I$ is free of rank one as a left
$E$-module and, when it is, return a free generator $\alpha$ satisfying
\[
        I=E\alpha.
\]
In that case
\[
        q:=\alpha j\in H
\]
is a compatible integral pair, say $q=(S,F)$. We accept exactly when
both component matrices are unimodular,
\[
        S\in\GL_2(\mathbb Z),
        \qquad
        F\in\GL_5(\mathbb Z).
\]
Because $q\in H$, it already satisfies
\[
        FX_T=X_{T'}S,
        \qquad
        F\Lambda_i\subseteq\Lambda_i.
\]
The block form forced by the two sublattice conditions shows that if
$F$ is unimodular, then its diagonal blocks are unimodular and hence
the two inclusions are equalities. Lemma~7.1 therefore gives a tensor
isomorphism.

Finally, it is enough to test the one free generator returned by the
algorithm. If an integral diagram isomorphism $q_0$ exists, then
\[
        H=Eq_0
\]
and hence
\[
        I=E(q_0j^{-1}).
\]
Thus $I$ is free of rank one. Any two free generators of $I$ differ
by multiplication by a unit of $E$. Such a unit is an integral
automorphism of the target diagram and has unimodular component
matrices. Therefore every free generator passes the final
unimodularity test whenever one does.

All lattices and rational normalizations above have fixed dimension
and are computable by polynomial-time integer and rational linear
algebra. The only quantum step is the cited
Bley-Hofmann-Johnston principality computation. Hence integral
isomorphism in the linear-kernel case is decidable in \BQP.
\end{proof}

\subsection{A quadratic kernel gives two coefficient matrices}
Suppose instead that $v_T(x,y)=x^2v_0+xyv_1+y^2v_2$ is the primitive least-degree kernel.
Write
\begin{equation}
K_T=\begin{pmatrix}v_0&v_1&v_2\end{pmatrix},\qquad
\nu(x,y)=\begin{pmatrix}x^2\\xy\\y^2\end{pmatrix},\qquad
L(x,y)=\begin{pmatrix}-y&x&0\\0&-y&x\end{pmatrix}.
\label{eq:quadratic-kernel-matrices}
\end{equation}
Here $K_T$ is a coefficient matrix, not the unused-direction lattice $K$ from~\eqref{eq:KH}.

\begin{lemma}\label{lem:quadratic-kernel-data}
The matrix $K_T$ is nonsingular. The saturated output span has rank two. After choosing an
integral basis of that span and deleting the unused output direction, the resulting
$2\times3$ pencil $\overline M_T$ determines a nonsingular $H_T\in\Mat_2(\Z)$ by
\begin{equation}
\overline M_T(x,y)K_T=H_TL(x,y).
\label{eq:quadratic-kernel-factorization}
\end{equation}
These data are computable in deterministic polynomial time.
\end{lemma}
\begin{proof}
If the coefficient vectors spanned a line, canceling their common polynomial factor would
give a constant kernel vector. If they spanned a plane, a rational coordinate change would
make the kernel vector $(f,g,0)^{\transpose}$, where $f,g$ are relatively prime homogeneous
quadratics: any common factor would give a lower-degree kernel. A row of the pencil then
gives $af+bg=0$ with $a,b$ linear. Relative primality forces $a=b=0$, since a quadratic
cannot divide a nonzero linear polynomial. Indeed, from \(af+bg=0\) and \(\gcd(f,g)=1\), we have
\(f\mid b\) and \(g\mid a\). Since \(f\) and \(g\) have degree two,
whereas \(a\) and \(b\) are linear forms, this forces \(a=b=0\).
Applying this to every row of the pencil shows that, in these
coordinates, only the third column of \(M_T(x,y)\) can be nonzero.
Hence the pencil would have normal rank at most one, contradicting
our standing assumption that its normal rank is two. The pencil would have at most one nonzero
column, contradicting normal rank two. Thus $K_T$ is nonsingular.

The linear polynomial relations among $x^2,xy,y^2$ are exactly the constant linear
combinations of the two rows of $L(x,y)$. To see this explicitly, suppose that linear forms \(a,b,c\) satisfy
\[
    ax^2+bxy+cy^2=0.
\]
Writing
\[
    a=a_1x+a_2y,\qquad
    b=b_1x+b_2y,\qquad
    c=c_1x+c_2y
\]
and comparing the coefficients of \(x^3,x^2y,xy^2,y^3\) gives
\[
    a_1=0,\qquad a_2+b_1=0,\qquad
    b_2+c_1=0,\qquad c_2=0.
\]
Thus \((a,b,c)\) is a constant linear combination of
\[
    (-y,x,0)
    \qquad\text{and}\qquad
    (0,-y,x),
\]
the two rows of \(L(x,y)\). Consequently every row of
\(M_T(x,y)K_T\) is such a combination, which is precisely why there
is a constant matrix \(\widetilde H_T\) with
\[
    M_T(x,y)K_T=\widetilde H_TL(x,y).
\] 
This follows immediately by comparing the four
cubic coefficients. Since $M_TK_T\nu=0$, we obtain $M_TK_T=\widetilde H_TL$ with
$\widetilde H_T\in\Mat_{3\times2}(\Z)$. Its entries are read directly from the linear
coefficients, and its rank is two because $M_T$ has normal rank two. The columns of the
two coefficient matrices of $L$ span $\Q^2$, so the total output span is
$\im\widetilde H_T$, of dimension two. Choose a basis of its saturation and apply
Lemma~\ref{lem:remove-unused}. The same identity becomes
\eqref{eq:quadratic-kernel-factorization}, with $H_T$ integral and nonsingular.
\end{proof}

For $S=\left(\begin{smallmatrix}r&s\\t&u\end{smallmatrix}\right)$, define
\begin{equation}
\Sigma_2(S)=\begin{pmatrix}r^2&2rs&s^2\\rt&ru+st&su\\t^2&2tu&u^2\end{pmatrix}.
\label{eq:kernel-symmetric-square}
\end{equation}
The purpose of \(\Sigma_2(S)\) is simply to record how a change of variables in
\((x,y)\) acts on the three quadratic monomials appearing in the kernel vector.
Indeed, if \(z=(x,y)^T\), then
\[
    \nu(Sz)=\Sigma_2(S)\nu(z).
\]
Thus a first-input change \(S\) on the original pencil induces the
\(3\times 3\) change \(\Sigma_2(S)\) on the coefficient vector
\((x^2,xy,y^2)^T\); no additional choice is being made here.  Since the rows
of \(L(z)\) generate all linear relations among these three monomials, the
relation matrix must transform compatibly as well.  Direct multiplication
gives
\begin{equation}
    L(Sz)\Sigma_2(S)=(\det S)S L(z),
    \qquad
    \det\Sigma_2(S)=(\det S)^3 .
\label{eq:quadratic-kernel-identity}
\end{equation}

\begin{proposition}\label{prop:symmetric-criterion}
In the quadratic-kernel case, $|\det K_T|$ and $|\det H_T|$ are isomorphism invariants.
After comparing these invariants, the tensors are isomorphic if and only if some
$S\in\GL_2(\Z)$ satisfies
\begin{equation}
H_{T'}SH_T^{-1}\in\Mat_2(\Z),\qquad
K_{T'}\Sigma_2(S)K_T^{-1}\in\Mat_3(\Z).
\label{eq:symmetric-integral-criterion}
\end{equation}
\end{proposition}
\begin{proof}
By Lemma~\ref{lem:remove-unused}, compare the reduced pencils. An isomorphism satisfies
$P\overline M_T(z)=\overline M_{T'}(Sz)Q$ with $P\in\GL_2(\Z)$ and $Q\in\GL_3(\Z)$.
It carries the primitive polynomial kernel to the target primitive polynomial kernel up
to sign. Indeed, both $Q$ and the coefficient substitution $\Sigma_2(S)$ are unimodular
integer maps, so they preserve the gcd of all coefficients. Absorb the sign by replacing
$P,Q$ simultaneously by their negatives. The input map is then forced to be
\[
Q=K_{T'}\Sigma_2(S)K_T^{-1}.
\]
Using~\eqref{eq:quadratic-kernel-factorization} and~\eqref{eq:quadratic-kernel-identity}
forces the output map to be
\[
P=(\det S)H_{T'}SH_T^{-1}.
\]
Taking determinants proves invariance of both absolute determinants. Conversely, after
these invariants match, the two displayed maps have determinant of absolute value one.
The integrality conditions~\eqref{eq:symmetric-integral-criterion} therefore make them
unimodular, and the same identities give an isomorphism. Either choice of sign of each
primitive kernel gives the same criterion.
\end{proof}
\section{Direct comparison by polynomial congruences}\label{sec:symmetric-comparison}
We now decide the quadratic-kernel criterion of Proposition~7.4.
At the end of Section~7, all data associated with the two tensors have
already been computed except for the first-factor change of variables
\[
        S=
        \begin{pmatrix}
        r & s\\
        t & u
        \end{pmatrix}
        \in \GL_2(\mathbb Z).
\]
Once $S$ is fixed, the changes of basis in the other two factors are forced.
Proposition~7.4 shows that tensor isomorphism is equivalent to the existence
of such an $S$ for which
\[
        H_{T'} S H_T^{-1}\in \Mat_2(\mathbb Z),
        \qquad
        K_{T'}\Sigma_2(S)K_T^{-1}\in \Mat_3(\mathbb Z).
\]
Thus we are not independently solving for the entries of three matrices.
The only unknowns are the four integers $r,s,t,u$. The entries of the two
displayed matrices are rational expressions determined by these four
unknowns.  The first matrix depends linearly on $r,s,t,u$, while the second
depends quadratically through $\Sigma_2(S)$.

The remaining difficulty is therefore an integrality problem rather than a
system of exact polynomial equations over $\mathbb Z$.  Choose a positive
integer $N$ clearing the fixed denominators of $H_T^{-1}$ and $K_T^{-1}$.
After multiplying by $N$, the two matrix conditions above are equivalent to
requiring certain integer polynomials in $r,s,t,u$ to be divisible by $N$.
The $2\times2$ matrix contributes four such congruences and the $3\times3$
matrix contributes nine, giving thirteen congruences in total.  In addition,
the requirement $S\in\GL_2(\mathbb Z)$ means
\[
        ru-st=\delta,
        \qquad \delta\in\{\pm1\},
\]
which gives one further quadratic congruence modulo $N$.

Consequently, for each fixed $\delta\in\{\pm1\}$, the remaining question has
the form
\[
        \exists\, r,s,t,u\in \mathbb Z/N\mathbb Z
        \quad
        \text{such that}
        \quad
        f_1(r,s,t,u)=\cdots=f_{14}(r,s,t,u)=0
        \quad\text{in }\mathbb Z/N\mathbb Z,
\]
where the $f_i$ are integer polynomials of degree at most two.  In this
sense, we seek a \emph{common zero}. One residue-class quadruple
$(r,s,t,u)$ must satisfy all of the congruences simultaneously. It is also
why quantifiers naturally enter the argument. The statement that such a
quadruple exists is an existential formula over the finite ring
$\mathbb Z/N\mathbb Z$.

Lemma~8.1 supplies the algorithmic tool for deciding exactly this fixed-format
existence problem once the modulus is factored.  Theorem~8.2 will then use
unimodular lifting to pass from a residue-class solution modulo $N$ to an
actual matrix $S\in\GL_2(\mathbb Z)$, and Proposition~7.4 will convert that
lift back into a tensor isomorphism.

\subsection{A fixed-format congruence algorithm}
The following algorithmic consequence of quantifier elimination is used with fixed numbers
of variables and equations and fixed degree. The coefficients and the modulus remain part
of the input; in particular, prime-power exponents are not bounded.

\begin{lemma}\label{lem:fixed-format-congruences}
Fix positive integers $n,m,d$. Let $f_1,\ldots,f_m$ be integer polynomials in $n$ variables
of degree at most $d$. Given their coefficients and an integer $N\ge2$ together with its prime
factorization, one can decide in deterministic polynomial time in the binary input length
whether the $f_i$ have a common zero over $\Z/N\Z$.
\end{lemma}
\begin{proof}
Write the given factorization as
\[
    N=\prod_{\nu=1}^{s}p_\nu^{e_\nu}.
\]
By the Chinese remainder theorem, the polynomials have a common zero
modulo $N$ if and only if they have a common zero modulo every
$p_\nu^{e_\nu}$.  It therefore suffices to give a polynomial-time
test for one prime power
\[
    q=p^e.
\]
The prime $p$ and exponent $e$ are part of the input and need not
be bounded.

Let $x=(x_1,\ldots,x_n)$.  For each monomial
$x^\alpha=x_1^{\alpha_1}\cdots x_n^{\alpha_n}$ of total degree
$|\alpha|=\alpha_1+\cdots+\alpha_n\le d$, introduce a coefficient
parameter $a_{i,\alpha}$, and put
\[
    F_i(a,x)=\sum_{|\alpha|\le d}a_{i,\alpha}x^\alpha,
    \qquad 1\le i\le m.
\]
Here $a$ denotes the tuple of all these coefficient parameters.
For a particular input, we substitute the coefficients of $f_i$
for the corresponding entries of $a$, using zero for absent
monomials.  Thus $F_i(a,x)=f_i(x)$ after specialization.
The entries of $a$ are given data. The entries of $x$ are the
unknowns we seek. Since $n,m,d$ are fixed, the number of
coefficient slots and the expressions $F_i$ are fixed, although
the input coefficients can have arbitrarily large bit length.

Let $\mathbb Q_p$ be the field of $p$-adic numbers, let $v_p$
be its valuation, normalized by $v_p(p)=1$ and
$v_p(0)=\infty$, and let
\[
    \mathbb Z_p=\{z\in\mathbb Q_p:v_p(z)\ge0\}.
\]
For a nonzero ordinary integer, $v_p$ counts its factors of $p$. The convention at zero reflects that zero is divisible by every
power of $p$.  More generally, for $z\in\mathbb Z_p$,
\[
    z\in p^e\mathbb Z_p
    \quad\Longleftrightarrow\quad
    v_p(z)\ge e=v_p(q).
\]
Since
$\mathbb Z_p/p^e\mathbb Z_p\cong\mathbb Z/p^e\mathbb Z$
and polynomial evaluation commutes with reduction, solvability
modulo $q$ is equivalent to the formula
\[
\begin{aligned}
    \Phi(a,q):={}&
    \exists x_1,\ldots,x_n\in\mathbb Q_p\\
    &\left(
       \bigwedge_{j=1}^{n} v_p(x_j)\ge0
       \;\wedge\;
       \bigwedge_{i=1}^{m}
          v_p\!\left(F_i(a,x)\right)\ge v_p(q)
     \right).
\end{aligned}
\]
Indeed, integer representatives of a residue solution give
witnesses for this formula, and reducing any witnesses modulo
$p^e$ gives a residue solution.  We are not asking for an exact
common zero in $\mathbb Z_p$, only for the polynomial values to
be divisible by $p^e$. This is one fixed formula with free parameters $a,q$.
Pas's language has three sorts: the valued field, its residue
field, and its value group
\cite[Definition~2.3]{pas}.
In our application these are
\[
    \mathbb Q_p,\qquad
    \mathbb F_p,\qquad
    \mathbb Z\cup\{\infty\},
\]
respectively.  Thus Pas's symbols $K$, $\overline K$, and
$\Gamma$ refer here to $\mathbb Q_p$, $\mathbb F_p$, and
$\mathbb Z$, with $\infty$ adjoined to the last sort.
His valuation symbol $\operatorname{ord}$ is our $v_p$.

Applying \cite[Corollary~4.3]{pas} to the fixed formula
$\Phi$ gives a fixed formula $\Psi(a,q)$ and a fixed finite
set of exceptional primes $\mathcal E$ such that, whenever
$p\notin\mathcal E$,
\[
    \Phi(a,q)\quad\Longleftrightarrow\quad\Psi(a,q)
\]
for every specialization of the free parameters.
The formula $\Psi$ has no quantified variables in the
valued-field sort.  In other words, the unknowns ranging over
$\mathbb Q_p$ have been eliminated.  Quantifiers over the
residue field and the value group may remain, and these will be
handled separately below.  Both $\Psi$ and $\mathcal E$ depend
only on the original fixed formula, hence only on $n,m,d$,
not on the input coefficients or on $e$.

The free valued-field parameters can occur in $\Psi$ only
through polynomial equalities and through valuations and
angular components of finitely many fixed polynomials
$h(a,q)$ where the angular component records the residue of a
number after its power of $p$ has been removed,
\[
    \operatorname{ac}_p(z)
       =p^{-v_p(z)}z\bmod p
       \quad(z\ne0),
    \qquad
    \operatorname{ac}_p(0)=0.
\]
Thus, if $z=p^r u$ with $v_p(u)=0$, its valuation is $r$
and its angular component is $u\bmod p$.
This is the angular-component map appearing in Pas's
language, and is the leading residue of $z$. 

It is useful to keep the roles of the residue-field and value-group sorts distinct.
For $\mathbb{Q}_p$, the value group is $\mathbb{Z}$ and the valuation
$v_p(z)$ records how many powers of $p$ occur in $z$. The residue field is
$\mathbb{F}_p=\mathbb{Z}_p/p\mathbb{Z}_p$: after removing this power of $p$,
the angular component $\operatorname{ac}_p(z)$ records the remaining unit
modulo $p$.  Thus the valuation and angular component retain two different
pieces of information about a $p$-adic number,
\[
        z=p^{v_p(z)}u
        \qquad\leadsto\qquad
        v_p(z)\in\mathbb{Z},
        \quad
        \operatorname{ac}_p(z)=u\bmod p\in\mathbb{F}_p .
\]
Pas's elimination removes the quantified variables ranging over the valued
field $\mathbb{Q}_p$, but it may leave quantified variables in these two
simpler sorts.  They are handled separately because their algebraic
structures are different. The value-group part is a formula in integer
addition, order, and congruences, hence belongs to Presburger arithmetic,
whereas the residue-field part is a formula in the ring language over the
finite field $\mathbb{F}_p$.  We therefore apply Cooper's quantifier
elimination to the former and the finite-field reduction of Kiefe to the
latter \cite{Coo72, Kie76}.

There are no maps between the residue-field and value-group
sorts in this language. Consequently, after a fixed Boolean
rearrangement, the remaining conditions separate into
residue-field formulas and value-group formulas, together
with polynomial zero tests.  Cases involving the value
$\infty$ are separated explicitly, including any quantified
value-group variable that equals $\infty$.

After the $\infty$ cases have been separated, the value-group
formulas are formulas of Presburger arithmetic, that is, arithmetic on
$\mathbb Z$ using addition, order, and congruences, but not
multiplication of two variables.  Cooper's quantifier-elimination
procedure \cite{Coo72} replaces them by fixed Boolean
combinations of integer linear comparisons and congruences
with fixed moduli. Their inputs are the computed valuations
$v_p(h(a,q))$. This elimination is performed on the fixed
formulas, before the numerical input parameters are substituted.

For the residue-field conditions, we use the following
consequence of Kiefe's theorem \cite{Kie76}; see also
\cite[Facts~4.3 and~4.5, arXiv version]{medvedev-vanabel}.
Every fixed formula in the language of rings is, over all
sufficiently large finite fields, equivalent to a Boolean
combination of formulas
\[
    \exists t:\ G(c,t)=0.
\]
Here $G(c,t)$ is one of the finitely many fixed auxiliary polynomials
produced by the Kiefe reduction. It is not one of the original input
polynomials $f_{i}$.  The tuple $c$ consists of the residue-field parameters
already obtained from the input, while $t$ is the single remaining
quantified residue-field variable. The equivalence holds for all choices of $c$, and the threshold depends only on the formula. In our application $c$ is a tuple of angular components already computed from the input. Enlarge $\mathcal E$ to include the finitely many primes excluded by these
residue-field reductions.

For $p\notin\mathcal E$, specializing $c$ therefore reduces
each residue-field condition to a fixed Boolean combination
of one-variable root tests for polynomials
\[
    g(t)\in\mathbb F_p[t]
\]
of fixed bounded degree.
The zero polynomial has a root, and a nonzero constant does
not.  For a nonconstant polynomial,
\[
    g \text{ has a root in }\mathbb F_p
    \quad\Longleftrightarrow\quad
    \deg\gcd\!\left(g(t),t^p-t\right)>0,
\]
because $t^p-t$ is the product of the distinct linear polynomials $t-b$ with $b\in\mathbb F_p$. We do not construct a polynomial of degree $p$. Instead, repeated squaring computes $t^p\bmod g$, after which a polynomial gcd computation decides the test. Since $\deg g$ is fixed, this takes deterministic time
polynomial in $\log p$.

For each $p\in\mathcal E$, apply Macintyre's fixed-$p$
quantifier elimination directly to the original solvability
condition \cite[Theorem~1]{macintyre}.
This condition is expressible in the field language:
integrality in $\mathbb Z_p$ is definable for fixed $p$,
and, since $q\ne0$,
\[
    v_p(F_i(a,x))\ge v_p(q)
    \quad\Longleftrightarrow\quad
    F_i(a,x)/q\in\mathbb Z_p.
\]
Macintyre's theorem therefore gives a fixed Boolean
combination of polynomial equalities and fixed-power
predicates on rational values obtained from $a,q$.

These predicates are also efficiently decidable.
For a nonzero rational value $b$, write
\[
    r=v_p(b),\qquad u=p^{-r}b,
\]
so that $v_p(u)=0$.
For each fixed positive integer $\ell$ occurring in a
power predicate,
\[
\begin{aligned}
    b\in(\mathbb Q_p^\times)^\ell
    \quad\Longleftrightarrow\quad&
    r\equiv0\pmod{\ell}
    \ \text{ and }\\
    &u\bmod p^{\,2v_p(\ell)+1}
      \text{ is an $\ell$-th power in }
      \bigl(\mathbb Z/p^{\,2v_p(\ell)+1}\mathbb Z\bigr)^\times .
\end{aligned}
\]
Necessity follows by taking valuations and reducing the unit
part of an $\ell$-th root.
For sufficiency, choose a unit representative $w$ satisfying the displayed congruence and apply
Hensel's lemma \cite[Chapter~II, \S2.2]{Ser73} to
$R(t)=t^\ell-u$. Indeed,
\[
    v_p(R(w))
       \ge 2v_p(\ell)+1
       >2v_p(R'(w)),
\]
because $w$ is a unit and
$v_p(R'(w))=v_p(\ell)$.
Thus $u$ has an $\ell$-th root in $\mathbb Z_p^\times$;
the divisibility of $r$ by $\ell$ then gives an $\ell$-th
root of $b$ in $\mathbb Q_p^\times$.
Zero values are treated separately.

Here both $p$ and every exponent $\ell$ are fixed constants,
so the finite residue test has constant size.  A rational
unit is reduced modulo this modulus by inverting its
denominator, which is prime to $p$.
The input exponent $e$ remains unrestricted.

For fixed $n,m,d$, all auxiliary polynomials, Boolean
combinations, power exponents, and exceptional-prime data
above are fixed. Evaluating a fixed polynomial at the input
integers or rational values produces numbers of polynomial
bit length. Their zero tests, valuations at the known
prime $p$, and required residues are computable in polynomial
time using ordinary integer arithmetic. The remaining
integer comparisons and finite-field root tests have the
same complexity.
\end{proof}

\subsection{Four variables and unimodular lifting}
Compute $K_T,H_T,K_{T'},H_{T'}$ as in Section~\ref{sec:kernel-comparison} and reject unless
the corresponding absolute determinants agree. Put
\begin{equation}
N=2\bigl|\det H_T\,\det K_T\bigr|.
\label{eq:sym-global-modulus}
\end{equation}
This clears the denominators of $H_T^{-1}$ and $K_T^{-1}$; the factor $2$ only ensures
$N\ge2$. For $S=\left(\begin{smallmatrix}r&s\\t&u\end{smallmatrix}\right)$, form
\begin{equation}
\mathcal P(S)=N H_{T'}SH_T^{-1},\qquad
\mathcal Q(S)=N K_{T'}\Sigma_2(S)K_T^{-1}.
\label{eq:direct-polynomial-matrices}
\end{equation}
These are polynomial matrices with integer coefficients, of degrees one and two,
respectively. They are formed over $\Q$ and multiplied by $N$ \emph{before} reducing
coefficients modulo $N$; no division is performed in the residue ring.
For each fixed $\delta\in\{\pm1\}$, test
\begin{equation}
\mathcal P(\bar S)\equiv0\pmod N,\qquad
\mathcal Q(\bar S)\equiv0\pmod N,\qquad
\det\bar S\equiv\delta\pmod N.
\label{eq:sym-finite-test}
\end{equation}
The matrix congruences are entrywise. Thus the test has four variables and fourteen
scalar equations, all of degree at most two.

\begin{theorem}\label{thm:symmetric-bqp}
For each $\delta\in\{\pm1\}$, the criterion~\eqref{eq:symmetric-integral-criterion} with
$\det S=\delta$ holds if and only if~\eqref{eq:sym-finite-test} is solvable.
Consequently the quadratic-kernel case is in \BQP, and is in deterministic classical
polynomial time given the factorization of $N$.
\end{theorem}
\begin{proof}
An integral solution reduces modulo $N$ to a solution of~\eqref{eq:sym-finite-test}.
Conversely, suppose $\bar S$ is a residue solution. Bley--Hofmann--Johnston's
unimodular lifting theorem~\cite[Corollary~7.6]{bhj} gives $S\in\GL_2(\Z)$ with
$S\equiv\bar S\pmod N$ and $\det S=\delta$. Multiply first by
$\operatorname{diag}(\delta,1)$, lift in $\SL_2$, and undo that multiplication.
Because $\mathcal P,\mathcal Q$ have integer coefficients, their values at $S$ are
divisible by $N$. The two matrices in~\eqref{eq:symmetric-integral-criterion} are
therefore integral. Their determinants have absolute value one by the preliminary
comparisons, so Proposition~\ref{prop:symmetric-criterion} supplies an isomorphism.

All constructed coefficients have polynomial bit length. Factor $N$ by Shor's
algorithm~\cite{shor} and apply Lemma~\ref{lem:fixed-format-congruences} to each sign.
The same sign is used at every prime-power factor. Accept exactly when one test succeeds.
Only existence is required from the congruence algorithm; the lift proves sufficiency
without requiring that algorithm to produce a residue solution.
\end{proof}

\begin{corollary}
Integral $2\times3\times3$ tensor isomorphism restricted to
$f_T\equiv0$ is in \BQP. The dependent-slice and
normal-rank-at-most-one cases are in deterministic polynomial time.
\end{corollary}
\begin{proof}
The elementary branches are ~\ref{lem:dependent-slices} and ~\ref{lem:low-normal-rank}. At normal rank two,
compare the ordered common-kernel dimensions. Two common kernels give
the known $2\times2\times2$ core algorithm. Otherwise transpose both
pencils when needed to arrange $K_R=0$. Lemma ~\ref{lem:polynomial-kernel} is exhaustive:
degree one uses Proposition ~\ref{prop:linear-kernel-bqp}, and degree two uses Proposition ~\ref{prop:symmetric-criterion}
and Theorem ~\ref{thm:symmetric-bqp}.
\end{proof}

\section{The complete algorithm and the arithmetic obstruction}
\begin{proof}[Proof of Theorem~\ref{thm:main}]
Compare the dimensions of the two slice spans. Unequal dimensions rule out
isomorphism. Dimensions zero and one are handled by Lemma~\ref{lem:dependent-slices}. For independent slices, compute
both determinant forms. The property of vanishing identically is invariant under isomorphism, so
reject if exactly one vanishes.

If both forms are nonzero, apply Corollary~\ref{cor:regular-locus}. More explicitly, compare their pure-cube status.
Non-pure cubes require at most six fixed-form tests, each performed by Proposition~\ref{prop:fixed-form}. Pure
cubes are normalized to $Cx^3$, after which the adapted flag calculation in Proposition~\ref{prop:pure-cube-test} decides
equivalence. Independence of the slices ensures that the normalized second slice is nonzero.

If both forms vanish, compare normal ranks using~\eqref{eq:normal-rank}. Rank at most one is covered by
Lemma~\ref{lem:low-normal-rank}. At rank two, compare the ordered dimensions of $K_R,K_L$. With two common kernels,
delete unused directions and compare the integral $2\times2\times2$ cores. Otherwise transpose both pencils
when needed to arrange $K_R=0$, and compute their least polynomial-kernel degrees using
Lemma~\ref{lem:polynomial-kernel}. Reject if the degrees differ. For degree one, form the compatible diagrams of
Lemma~\ref{lem:linear-kernel-diagram} and apply
Proposition~\ref{prop:linear-kernel-bqp}.
For degree two, compute the coefficient matrices $K_T,H_T$ and $K_{T'},H_{T'}$ from
Lemma~\ref{lem:quadratic-kernel-data}, and compare their absolute determinants. Apply the direct
congruence test of Theorem~\ref{thm:symmetric-bqp}.

These cases are exhaustive. The classical steps use rational linear algebra, fixed-degree polynomial
factorization, primitive lattice bases, Smith or Hermite normal form, lattice intersections and projections, and gcds.
The quantum arithmetic consists of the compatible-order tests, the known $2\times2\times2$ core algorithm,
and factoring the direct-comparison modulus~\eqref{eq:sym-global-modulus}. Given that factorization,
the fixed-format congruence test is deterministic polynomial time by Theorem~\ref{thm:symmetric-bqp}.
\end{proof}

\subsection{Faithful padding}
For $C=(C_1,C_2)\in\Z^2\tensor\Z^2\tensor\Z^2$, define
\begin{equation}
\iota(C)_i=
\begin{pmatrix}
C_i&0\\
0&0
\end{pmatrix}
\in\Mat_3(\Z),
\qquad i=1,2.
\label{eq:padding}
\end{equation}
Its determinant form vanishes identically.

\begin{proposition}\label{prop:faithful-padding}
For all integral $2\times2\times2$ tensors $C,C'$, one has
\[
\iota(C)\cong\iota(C')\qquad\Longleftrightarrow\qquad C\cong C'.
\]
\end{proposition}
\begin{proof}
An isomorphism of the smaller tensors extends across the added input and output directions.
Conversely, padding adds only unused directions. Removing all unused directions from $C$ or $\iota(C)$
gives the same reduced map~\eqref{eq:reduced-map}, and similarly for $C'$. An isomorphism of the padded tensors
therefore induces one of the reduced maps. Apply Lemma~\ref{lem:remove-unused} in the original $2\times2\times2$ dimensions
to extend it to an isomorphism of $C,C'$.
\end{proof}

\begin{corollary}\label{cor:principality-obstruction}
The real-quadratic principality decision problem considered by Grochow--Qiao--
Stange--Sun reduces in polynomial time to identically singular integral $2\times3\times3$ tensor isomorphism.
\end{corollary}
\begin{proof}
Compose their reduction to integral $2\times2\times2$ tensor isomorphism~\cite[Theorem~3.7]{gqss} with
Proposition~\ref{prop:faithful-padding}.
\end{proof}

The regular and zero-form loci thus fit one structural picture, while requiring different arithmetic
descriptions. For nonzero forms, Bhargava--Wood balanced modules retain the pairing through a
compatible diagram. On the boundary, a least-degree polynomial kernel retains the singular
structure: its linear case yields another compatible diagram, and its quadratic case yields two
coefficient matrices with compatible linear and quadratic transformations. These lead respectively
to a compatible-order test and a direct finite congruence test. In both cases the essential condition is not that the individual
lattices be isomorphic, but that their isomorphisms preserve the structure linking them.

\begingroup
\setlength{\intextsep}{6pt}
\begin{algorithm}[H]
\small
\caption{Integral $2\times3\times3$ tensor isomorphism}
\label{alg:integral-233-ti}
\begin{algorithmic}[1]
\Require $T=(A,B)$ and $T'=(A',B')$ with integral $3\times3$ slices.
\Ensure \textsc{Isomorphic} or \textsc{Not Isomorphic}, with error at most $1/3$.
\State Compare slice-span dimensions; reject if unequal.
\State If the slices are dependent, return the Smith-form test of Lemma~\ref{lem:dependent-slices}.
\State Compute $f=f_T$ and $f'=f_{T'}$; reject if exactly one is zero.
\If{$f,f'\not\equiv0$}
    \State Reject if exactly one form is a pure cube.
    \If{both are pure cubes}
        \State Normalize to $Cx^3$ and $C'x^3$; reject if $C\neq C'$.
        \State Form the adapted pairs of Lemma~\ref{lem:adapted-form}; reject if $(a,d,c)\neq(a',d',c')$.
        \State Return the sign-enumerated linear/congruence test of Proposition~\ref{prop:pure-cube-test}.
    \Else
        \For{each $h$ with $f^h=f'$ listed by Lemma~\ref{lem:first-factor}}
            \State Compare $h\cdot T$ with $T'$ by Proposition~\ref{prop:fixed-form}; accept if successful.
        \EndFor
        \State \Return \textsc{Not Isomorphic}.
    \EndIf
\EndIf
\State Compare normal ranks; reject if unequal.
\State If both are at most one, return the test of Lemma~\ref{lem:low-normal-rank}.
\State Compare the ordered common-kernel dimensions; reject if unequal.
\If{both common kernels are nonzero}
    \State Delete unused directions and return the $2\times2\times2$ core test.
\EndIf
\State If $K_R\neq0$, transpose both pencils, so that $K_R=0$.
\State Compute $d_T,d_{T'}$ by Lemma~\ref{lem:polynomial-kernel}; reject if unequal.
\If{$d_T=1$}
    \State Form the two compatible diagrams of Lemma~\ref{lem:linear-kernel-diagram}.
    \State Return the compatible-order test of Proposition~\ref{prop:linear-kernel-bqp}.
\Else
    \State Compute $K_T,H_T,K_{T'},H_{T'}$ by Lemma~\ref{lem:quadratic-kernel-data}.
    \State Reject unless their corresponding absolute determinants agree.
    \State Form $N,\mathcal P,\mathcal Q$ by~\eqref{eq:sym-global-modulus}--\eqref{eq:direct-polynomial-matrices}; factor $N$.
    \For{each $\delta\in\{\pm1\}$}
        \State Decide~\eqref{eq:sym-finite-test} modulo $N$ by Lemma~\ref{lem:fixed-format-congruences}; accept if solvable.
    \EndFor
    \State \Return \textsc{Not Isomorphic}.
\EndIf
\end{algorithmic}
\end{algorithm}
\endgroup
\clearpage

\section*{AI Statement}
The general proof strategy used throughout this paper was developed by the authors without the use of AI. AI was used in exploring and refining proofs for the case where the determinant form is identically zero, as well as in the development of the wrapper lemmas for the lattice isomorphism algorithms. Specifically, ChatGPT Astra was used to help find a proof of Lemma 8.1, was used to help refine the proofs of propositions 4.4 and 7.2, and was also used to help draft and revise the text as well as to improve the presentation of Figure 1. The author has verified all proofs, added significant exposition throughout the paper, and take full responsibility for all of the results that appear in this paper and their correctness.

\section*{Acknowledgments}
The author thanks Youming Qiao for many helpful discussions and for providing useful feedback.

\bibliographystyle{alpha}
\bibliography{references}

\appendix

\section{An alternative proof for nondegenerate tensors via Bhargava II}
\label{app:bhargava:nondegenerate}

This appendix gives the alternative argument for integral
$2\times3\times3$ tensor isomorphism on the nondegenerate locus
\[
    \operatorname{Disc}(f_T)\ne 0,
    \qquad
    f_T(x,y)=\det(xA-yB),
    \qquad T=(A,B).
\]
The argument uses the cubic rings and balanced pairs of fractional ideals
in Bhargava's \emph{Higher composition laws II} directly
\cite{bhargava}.  It applies to the nonzero-discriminant case, not to
all tensors with $f_T\not\equiv 0$.  The notation introduced below is
local to this appendix; in particular, $K$ denotes a rational cubic
algebra.

\subsection{Fractional ideals, orders, and balanced pairs}
\label{app:bhargava:ideal-notation}

Let $R$ be a nondegenerate cubic order and let
\[
    K=R\otimes_{\mathbb Z}\mathbb Q.
\]
Thus $K$ is an \'etale cubic $\mathbb Q$-algebra: it is a product of
number fields whose degrees sum to three.  Following the terminology
used in the quadratic setting of \cite{gqss}, a fractional ideal of $R$
is a full rank-three $\mathbb Z$-lattice $I\subset K$ that is stable
under multiplication by $R$, that is,
\[
    RI\subseteq I.
\]
For a fractional ideal $I$, define its order by
\[
    \operatorname{ord}(I)
       \mathrel{:=}\{\alpha\in K:\alpha I\subseteq I\}.
\]
Its unit group is
\[
    \operatorname{ord}(I)^\times
       =\{u\in K^\times:uI=I\}.
\]
As observed in \cite{gqss}, $\operatorname{ord}(I)$ is computable in
polynomial time from a basis of $I$ by linear algebra.

The norm of a fractional ideal is its covolume relative to $R$:
\[
    N(I)=\frac{\operatorname{covol}(I)}
                 {\operatorname{covol}(R)},
\]
where both covolumes are computed in the same rational coordinates
on $K$.  In Bhargava's terminology \cite{bhargava}, a pair
$(I,I')$ of fractional $R$-ideals is \emph{balanced} if
\[
    II'\subseteq R,
    \qquad N(I)N(I')=1.
\]
Two balanced pairs $(I_1,I_1')$ and $(I_2,I_2')$
are equivalent if there exists $\kappa\in K^\times$ such that
\[
    I_1=\kappa I_2,
    \qquad I_1'=\kappa^{-1}I_2'.
\]

For orders $S_1,S_2\subset K$, we use the product-subgroup notation
\[
    S_1^\times S_2^\times
       =\{u_1u_2:u_1\in S_1^\times,\ u_2\in S_2^\times\}
       \subseteq K^\times.
\]

\subsection{Bhargava's parametrization and the tensor action}
\label{app:bhargava:parametrization}

For the tensor action, write
\[
\begin{aligned}
    \Gamma
       &=\operatorname{GL}_2(\mathbb Z)
          \times\operatorname{GL}_3(\mathbb Z)
          \times\operatorname{GL}_3(\mathbb Z),\\
    \Gamma_0
       &=\operatorname{GL}_2(\mathbb Z)
          \times\operatorname{SL}_3(\mathbb Z)
          \times\operatorname{SL}_3(\mathbb Z).
\end{aligned}
\]
These act on $\mathbb Z^2\otimes\mathbb Z^3\otimes\mathbb Z^3$ with
the tensor convention of the main text.

\begin{lemma}[{\cite[Section~2.1]{bhargava}}]
\label{app:bhargava:equivalent-orbits}
The $\Gamma$-orbits and the $\Gamma_0$-orbits on
$\mathbb Z^2\otimes\mathbb Z^3\otimes\mathbb Z^3$ coincide.
\end{lemma}

We write $R(f)$ for the cubic ring associated with a binary cubic form
$f$, corresponding to the notation $R_f$ in the main text.  Bhargava's
form of the Delone--Faddeev correspondence uses the determinant-twisted
$\operatorname{GL}_2(\mathbb Z)$-action on binary cubic forms; its
formula is recorded in Subsection~\ref{app:bhargava:cubic-rings} below.

\begin{lemma}[{\cite[Theorem~1]{bhargava}}]
\label{app:bhargava:delone-faddeev}
There is a canonical bijection between
$\operatorname{GL}_2(\mathbb Z)$-equivalence classes of integral binary
cubic forms, for this determinant-twisted action, and isomorphism
classes of cubic rings.  Moreover,
\[
    \operatorname{Disc}(f)=\operatorname{Disc}(R(f)).
\]
\end{lemma}

\begin{theorem}[{\cite[Theorem~2]{bhargava}}]
\label{app:bhargava:equivalence}
There is a canonical bijection between
\begin{enumerate}
    \item nondegenerate $\Gamma_0$-orbits on
    $\mathbb Z^2\otimes\mathbb Z^3\otimes\mathbb Z^3$; and
    \item isomorphism classes of pairs $(R;(I,I'))$, where $R$ is a
    nondegenerate cubic ring and $(I,I')$ is an equivalence class of
    balanced pairs of fractional $R$-ideals.
\end{enumerate}
\end{theorem}

\subsection{Efficiency of Bhargava's correspondence}
\label{app:bhargava:effective-correspondence}

We use Bhargava's parametrization in its explicit form.  In the proof
of Theorem~2 of \cite{bhargava}, one can explicitly recover from a
box $T=(A,B)$ its associated cubic ring $R$ and a representative
balanced pair of fractional $R$-ideals $(I,I')$.  Conversely, the box
is recovered from bases of these ideals.  Since all dimensions are
fixed, these constructions involve only a constant number of exact
integer/rational linear-algebra operations and are polynomial-time
computable in the standard basis representation.

We therefore adopt the same algorithmic convention as in the
$2\times2\times2$ setting of \cite{gqss}, where the explicit maps in
Bhargava's parametrization are likewise treated as polynomially
computable.  Throughout the remainder of this appendix, we use the
efficiently computable passage
\[
    T\longmapsto\bigl(R;(I,I')\bigr).
\]

\subsection{Lattice isomorphism background}
\label{app:bhargava:lattice-isomorphism}

\begin{proposition}[{\cite{bhj}}]
\label{app:bhargava:ideal-isomorphism}
Let $R$ be a nondegenerate cubic order and let
\[
    K=R\otimes_{\mathbb Z}\mathbb Q.
\]
Given two full fractional $R$-ideals $I,J\subset K$, one can decide
in \BQP whether $I\cong J$.  In particular, principality of a full
fractional $R$-ideal is decidable in \BQP.
\end{proposition}

\begin{proof}
Use the decision part of \textsc{IsPrincipal} from
\cite[Algorithm~8.3 and Remark~8.8(b)]{bhj}.  Hypothesis~(H) is
automatic since $K$ is a commutative \'etale $\mathbb Q$-algebra:
every simple component is a matrix algebra of size one over a number
field.  The required decomposition is computable in deterministic
polynomial time; see
\cite[Theorem~1.9]{lenstra2016algorithmscommutativealgebrasrational}.
This gives a principality decision procedure for full fractional
ideals over orders in $K$.

For the isomorphism test, put $S=\operatorname{ord}(J)$ and
$C=\{a\in K:aJ\subseteq I\}$.  By
\cite[Proposition~3.2]{bhj}, $I\cong J$ if and only if $C$ is
principal over $S$ and $CJ=I$, where $CJ$ is the additive lattice
generated by the products $cj$.  The bases of $S$, $C$, and $CJ$,
and the equality test, are computable in polynomial time by integer
linear algebra from the input bases and multiplication table.  Thus
only a principality decision is needed; no generator or isomorphism
matrix is computed.
\end{proof}

\subsection{From lattice isomorphisms to scalars}
\label{app:bhargava:scalars}

An $R$-module isomorphism between full fractional $R$-ideals extends,
after tensoring with $\mathbb Q$, to a $K$-linear automorphism of $K$.
Since $K$ is free of rank one as a module over itself, every such
automorphism is multiplication by a unique element
$\kappa\in K^\times$.  Thus, if $J\cong I$, then
\[
    I=\kappa J
\]
for some $\kappa\in K^\times$.  Here uniqueness refers to the scalar
representing a \emph{specified isomorphism}; the two ideals alone do
not determine $\kappa$ uniquely.

We use this statement only as an algebraic characterization.  The
decision algorithm below works directly with lattice bases and does
not compute $\kappa$, whose expanded coordinates need not have
polynomial bit length.

\subsection{Cubic-ring equivalence}
\label{app:bhargava:cubic-rings}

\begin{lemma}
\label{app:bhargava:cubic-ring-isomorphisms}
Let
\[
\begin{aligned}
    f  &=aX^3+bX^2Y+cXY^2+dY^3,\\
    f' &=a'X^3+b'X^2Y+c'XY^2+d'Y^3
\end{aligned}
\]
be nondegenerate integral binary cubic forms, and let
$R=R(f)$ and $R'=R(f')$ be their associated cubic rings.  There is a
deterministic polynomial-time algorithm that enumerates the at most
six ring isomorphisms
\[
    \phi:R\longrightarrow R'.
\]
\end{lemma}

\begin{proof}
By the Delone--Faddeev correspondence, in the form used by
\cite{bhargava}, isomorphism classes of cubic rings are identified
with $\operatorname{GL}_2(\mathbb Z)$-equivalence classes of integral
binary cubic forms.  For
\[
    M=\begin{pmatrix}r&s\\t&u\end{pmatrix}
      \in\operatorname{GL}_2(\mathbb Z),
\]
write
\[
    f^M(X,Y)=\det(M)^{-1}f(rX+tY,sX+uY).
\]
Unlike the untwisted substitution notation $f^h$ in the main text,
$f^M$ in this subsection includes the factor $\det(M)^{-1}$.
The equivalence algorithm of
\cite{cremona} enumerates in
deterministic polynomial time all matrices
$M\in\operatorname{GL}_2(\mathbb Z)$ satisfying $f'=f^M$, and there
are at most six possibilities.

For each such matrix $M$, the explicit Delone--Faddeev basis-change
construction gives the corresponding isomorphism between the cubic
rings $R(f)$ and $R(f')$.  This conversion involves only
fixed-dimensional integer arithmetic and is therefore deterministic
polynomial time.  Conversely, every ring isomorphism
$R(f)\to R(f')$ arises from one of these
$\operatorname{GL}_2(\mathbb Z)$-equivalences by the Delone--Faddeev
correspondence.  Hence the resulting list is complete and has size
at most six.
\end{proof}

\subsection{Balanced-pair equivalence}
\label{app:bhargava:balanced-pairs}

\begin{definition}[Balanced-pair equivalence decision problem]
Given balanced pairs of full fractional $R$-ideals $(I_1,I_2)$ and
$(J_1,J_2)$, decide whether there exists $\kappa\in K^\times$ such
that
\[
    I_1=\kappa J_1,
    \qquad I_2=\kappa^{-1}J_2.
\]
\end{definition}

\begin{lemma}
\label{app:bhargava:balanced-coupling}
Suppose that $I_1\cong J_1$ and $I_2\cong J_2$.  Choose
$\kappa_1,\kappa_2\in K^\times$ satisfying
$I_1=\kappa_1J_1$ and $I_2=\kappa_2J_2$, and set
\[
    S_1=\operatorname{ord}(J_1),
    \qquad S_2=\operatorname{ord}(J_2).
\]
Then the pair $(I_1,I_2)$ is equivalent to $(J_1,J_2)$ if and only if
\[
    (\kappa_1\kappa_2)^{-1}\in S_1^\times S_2^\times.
\]
\end{lemma}

\begin{proof}
Every scalar taking $J_1$ to $I_1$ has the form $\kappa_1u_1$ for
some $u_1\in S_1^\times$, since $\kappa J_1=\kappa_1J_1$ if and
only if $(\kappa_1^{-1}\kappa)J_1=J_1$, which is true if and only
if $\kappa_1^{-1}\kappa\in S_1^\times$.  The same argument applies
to every scalar taking $J_2$ to $I_2$.  The equivalence of
\cite{bhargava} requires that the second scalar is the inverse
of the first, so that
\[
    \kappa_2u_2=(\kappa_1u_1)^{-1}.
\]
Since $K$ is commutative, this is equivalent to
\[
    u_1u_2=(\kappa_1\kappa_2)^{-1},
\]
which is the desired membership condition.
\end{proof}

\subsection{A decision criterion using compatible scalar lattices}
\label{app:bhargava:compatible-scalars}

For fractional lattices $U,V\subset K$, let $UV$ denote the additive
lattice generated by the products $uv$.  The decision algorithm
encodes the two required scalar containments in a single lattice.

\begin{lemma}[Compatible scalar criterion]
\label{app:bhargava:compatible-scalar-criterion}
Let $(I_1,I_2)$ and $(J_1,J_2)$ be balanced pairs of full fractional
$R$-ideals.  Define
\[
\begin{aligned}
    S&=\operatorname{ord}(J_1)\cap\operatorname{ord}(I_2),\\
    L&=\{a\in K:aJ_1\subseteq I_1,\ aI_2\subseteq J_2\}.
\end{aligned}
\]
Then $S$ is an order in $K$, $L$ is a full fractional $S$-ideal,
and the pairs are equivalent if and only if
\[
    L\text{ is principal over }S,
    \qquad LJ_1=I_1,
    \qquad LI_2=J_2.
\]
\end{lemma}

\begin{proof}
The two orders defining $S$ contain $R$, so their intersection is
a full order.  Clearing denominators in multiplication matrices
shows that each containment defining $L$ gives a full lattice in
$K$, hence so does their intersection.  If $a\in L$ and $s\in S$,
then $(as)J_1\subseteq aJ_1\subseteq I_1$ and
$(as)I_2\subseteq aI_2\subseteq J_2$.  Thus $L$ is an $S$-ideal.

Suppose the pairs are equivalent via $\kappa\in K^\times$, so that
$\kappa J_1=I_1$ and $\kappa I_2=J_2$.  For $a\in K$, the
condition $a\in L$ is then equivalent to
$(\kappa^{-1}a)J_1\subseteq J_1$ and
$(\kappa^{-1}a)I_2\subseteq I_2$.  Therefore $L=\kappa S$.
Since $S$ contains $1$ and stabilizes both source lattices,
$SJ_1=J_1$ and $SI_2=I_2$.  Hence $LJ_1=I_1$ and $LI_2=J_2$.

Conversely, suppose the three stated conditions hold and write
$L=S\alpha$.  Fullness gives $K=K\alpha$, so
$\alpha\in K^\times$.  Commutativity and the identities
$SJ_1=J_1$, $SI_2=I_2$ give
\[
    I_1=LJ_1=\alpha SJ_1=\alpha J_1,
    \qquad J_2=LI_2=\alpha SI_2=\alpha I_2.
\]
Thus the pairs are equivalent.  The element $\alpha$ is used only
in this proof, not in the decision procedure.
\end{proof}

\begin{theorem}
\label{app:bhargava:balanced-pair-bqp}
Let $R$ be a nondegenerate cubic order, and let $(I_1,I_2)$ and
$(J_1,J_2)$ be balanced pairs of fractional $R$-ideals.  Deciding the
equivalence of the pairs $(I_1,I_2)$ and $(J_1,J_2)$ is in \BQP.
\end{theorem}

\begin{proof}
Compute $S$ and $L$ from
Lemma~\ref{app:bhargava:compatible-scalar-criterion}.  In rational
coordinates on $K$, multiplication by an unknown $a$ is linear in
its three coordinates.  The containments defining $S$ and $L$ are
therefore integral-coordinate conditions with rational coefficients
of polynomial bit length.  Clearing denominators and using Smith
or Hermite normal form computes bases of these full lattices in
polynomial time and with polynomial bit length.

Each of $LJ_1$ and $LI_2$ is generated by the nine pairwise products
of the respective lattice basis elements.  Their coordinates have
polynomial bit length, and equality with $I_1$ and $J_2$ is tested
by Hermite normal form.  Finally, use the principality decision in
Proposition~\ref{app:bhargava:ideal-isomorphism} for the $S$-ideal
$L$.  If an integral ideal is required, choose a nonzero integer
$d$ of polynomial bit length with $dL\subseteq S$ and test $dL$;
this does not change principality.  Accept exactly when both lattice
equalities and the principality test hold.  The preceding lemma
proves correctness, and no ideal generator or scalar witness is
computed.
\end{proof}

\subsection{Proof of the nondegenerate theorem}
\label{app:bhargava:main-proof}

\begin{theorem}
\label{app:bhargava:nondegenerate-bqp}
Deciding whether two nondegenerate integral $2\times3\times3$
tensors are isomorphic under $\Gamma$ is computable in \BQP.
\end{theorem}

\begin{proof}
Let $(R;(I_1,I_2))$ be the arithmetic data associated with $T$ and
$(R';(J_1,J_2))$ the arithmetic data associated with $T'$.  Write
\[
    K=R\otimes_{\mathbb Z}\mathbb Q,
    \qquad K'=R'\otimes_{\mathbb Z}\mathbb Q.
\]
By Lemma~\ref{app:bhargava:equivalent-orbits} and
Theorem~\ref{app:bhargava:equivalence}, two nondegenerate tensors
are isomorphic under
$\operatorname{GL}_2(\mathbb Z)\times\operatorname{GL}_3(\mathbb Z)
\times\operatorname{GL}_3(\mathbb Z)$ if and only if their associated
cubic rings and balanced ideal pairs are isomorphic.

Use Lemma~\ref{app:bhargava:cubic-ring-isomorphisms} to enumerate
the at most six ring isomorphisms $\phi:R\to R'$.  Each such
isomorphism extends uniquely to a $\mathbb Q$-algebra isomorphism
$K\to K'$, which we also denote by $\phi$.  For each $\phi$, transport
the balanced ideal pair $(I_1,I_2)$ to $(\phi(I_1),\phi(I_2))$ in
$R'$ and apply Theorem~\ref{app:bhargava:balanced-pair-bqp} to check
its equivalence with $(J_1,J_2)$.  Accept if and only if one of the
computed $\phi$ succeeds.
\end{proof}

\begin{algorithm}[H]
\caption{Nondegenerate integral $2\times3\times3$ tensor isomorphism
via Bhargava II}
\label{app:bhargava:algorithm}
\begin{enumerate}
    \item Use the construction of \cite{bhargava} to compute
    \[
        T\longmapsto(R;(I_1,I_2)),
        \qquad T'\longmapsto(R';(J_1,J_2)).
    \]
    \item Enumerate all $\phi:R\to R'$.  If none exists, return
    \textsc{NotIsomorphic}.
    \item For each $\phi$:
    \begin{enumerate}
        \item Set $\widetilde I_i=\phi(I_i)$ for $i=1,2$.
        \item Compute
        \[
        \begin{aligned}
            S&=\operatorname{ord}(J_1)
                \cap\operatorname{ord}(\widetilde I_2),\\
            L&=\{a\in K':aJ_1\subseteq\widetilde I_1,
                \ a\widetilde I_2\subseteq J_2\}.
        \end{aligned}
        \]
        \item If $LJ_1\ne\widetilde I_1$ or
        $L\widetilde I_2\ne J_2$, continue to the next $\phi$.
        \item Use the BHJ principality decision procedure for $L$
        over $S$.  If it succeeds, return \textsc{Isomorphic}.
    \end{enumerate}
    \item Return \textsc{NotIsomorphic}.
\end{enumerate}
\end{algorithm}

\end{document}